\documentclass[12pt,twoside,leqno]{article}
\usepackage{epsfig}
\usepackage{amsmath}
\usepackage{amsthm}
\usepackage{ifpdf}
\begin{document}

\theoremstyle{remark}
\newtheorem{remark}{Remark}

\parskip 4pt
\abovedisplayskip 7pt
\belowdisplayskip 7pt

\parindent=12pt

\newcommand{\A}{{\bf A}}
\newcommand{\B}{{\bf B}}
\newcommand{\bco}{{\boldsymbol{:}}}
\newcommand{\blambda}{{\boldsymbol{\lambda}}}
\newcommand{\bmu}{{\boldsymbol{\mu}}}
\newcommand{\bn}{{\bf n}}
\newcommand{\bnabla}{{\boldsymbol{\nabla}}}
\newcommand{\bomega}{{\boldsymbol{\omega}}}
\newcommand{\bsigma}{{\boldsymbol{\sigma}}}
\newcommand{\btheta}{{\boldsymbol{\theta}}}
\newcommand{\bg}{{\bf g}}
\newcommand{\bu}{{\bf u}}
\newcommand{\bU}{{\bf U}}
\newcommand{\bv}{{\bf v}}
\newcommand{\bw}{{\bf w}}
\newcommand{\bzero}{{\bf 0}}
\newcommand{\ct}{{\mathcal{T}}}
\newcommand{\cth}{{\mathcal{T}_h}}
\newcommand{\dsum}{{\displaystyle\sum}}
\newcommand{\D}{{\bf D}}
\newcommand{\e}{{\bf e}}
\newcommand{\F}{{\bf F}}
\newcommand{\G}{{\bf G}}
\newcommand{\g}{{\bf g}}
\newcommand{\Gx}{{{\overrightarrow{\bf Gx}}^{\perp}}}
\newcommand{\I}{{\bf I}}
\newcommand{\intbt}{{\displaystyle{\int_{B(t)}}}}
\newcommand{\intG}{{\displaystyle{\int_{\Gamma}}}}
\newcommand{\into}{{\displaystyle{\int_{\Omega}}}}
\newcommand{\intpb}{{\displaystyle{\int_{\partial B}}}}
\newcommand{\lto}{{L^2(\Omega)}}
\newcommand{\no}{{\noindent}}
\newcommand{\obo}{{\Omega \backslash \overline{B(0)}}}
\newcommand{\obt}{{\Omega \backslash \overline{B(t)}}}
\newcommand{\oo}{{\overline{\Omega}}}
\newcommand{\R}{{\text{I\!R}}}
\newcommand{\T}{{\bf T}}
\newcommand{\V}{{\bf V}}
\newcommand{\w}{{\bf w}}
\newcommand{\x}{{\bf x}}
\newcommand{\Y}{{\bf Y}}
\newcommand{\y}{{\bf y}}

\newpage
\thispagestyle{empty}

\noindent{\Large\bf A numerical study of the motion of a neutrally \\buoyant 
cylinder in two dimensional shear flow}

\bigskip
\normalsize \noindent{Tsorng-Whay Pan$^{a,}$\footnote{Corresponding author: e-mail:  pan@math.uh.edu}, Shih-Lin Huang$^b$, Shih-Di Chen$^b$, Chin-Chou Chu$^b$, \\Chien-Cheng Chang$^{b,c}$} \vskip 1ex
\noindent{$^a$\small Department of Mathematics, University of Houston, Houston,
Texas  77204, USA} \vskip 1ex
\noindent{$^b$\small Institute of Applied Mechanics, National Taiwan University, Taipei 106, Taiwan, ROC} \vskip 1ex
\noindent{$^c$\small Department of Mathematics and Taida Institute of Mathematical Sciences, National Taiwan University, Taipei 106, Taiwan, ROC}

\vskip 2ex
\noindent {\bf Abstract}
\vskip 1ex
In this paper, we investigate the motion of a neutrally
buoyant cylinder of circular or elliptic shape in two dimensional shear flow 
of a Newtonian fluid by direct numerical simulation. 
The numerical results are validated by comparisons with 
existing theoretical, experimental and numerical results, including a power law of the 
normalized angular speed versus the particle Reynolds number. 
The centerline between two walls is an expected equilibrium position of 
the cylinder mass center in shear flow. When placing the particle away 
from the centerline initially, it migrates toward another equilibrium 
position for higher Reynolds numbers due to the interplay between 
the slip velocity, the Magnus force, and the wall repulsion force.
\vskip 1ex
\noindent{\it keywords:} Shear flow; Neutrally buoyant particle; Equilibrium height;
Fictitious domain/distributed Lagrange multiplier method; Finite element.

\vskip 10ex
\noindent{\large\bf 1. Introduction}
\vskip 2ex

The problem of particle motions in shear flows is crucially important in many
engineering fields such as the handling of a fluid-solid mixture in slurry, colloid,
and fluidized bed. Segr\'e and Silberberg \cite{Segre1961, Segre1962} experimentally studied the 
migration of dilute suspensions of neutrally buoyant spheres in a tube Poiseuille flow. 
The particles migrate away from the wall and the centerline to accumulate at about 0.6 
of the tube radius from the centerline.  The experiments of Segr\'e and Silberberg 
\cite{Segre1961, Segre1962}  have had a large influence on fluid mechanics studies of migration 
and lift of particles. Comprehensive reviews of experimental and theoretical works 
have been given by Brenner \cite{Brenner1966}, Cox and Mason \cite{Cox1971},  Feuillebois \cite{Feuillebois1989}
and Leal \cite{Leal1980}.

Among the theoretical studies of the neutrally buoyant particle migration in linear shear flow,  
Bretherton \cite{Bretherton1962} found an expression for the lift force per unit length on a 
cylinder in an unbounded two-dimensional linear shear flow at small Reynolds number. 
Saffman's lift force \cite{Saffman1965} on a sphere 
of radius $a$ in an unbounded linear shear flow with shear rate $G$ is 
$F_s=6.46\rho V a^2 (G\nu)^{1/2}=6.46\rho \nu a V (Re_p)^{1/2}$ where 
$\nu$ is the kinetic viscosity of the fluid, 
$\rho$ is the density of the fluid, $Re_p=G a^2/\nu$ is the particle Renolds number, 
and $V$ is the slip velocity of the sphere.  In a bounded linear shear flow,
Ho and Leal \cite{Ho1974} examined the motion of a rigid sphere  with inclusion of the inertia effects at small Reynolds numbers by a regular perturbation method. The sphere reaches a stable lateral equilibrium position which is the midway between the walls. 
Vasseur and Cox \cite{Vasseur1976} also obtained the same stable lateral equilibrium position.
Ho and Leal require that $Re_p/\kappa^2 \ll 1$ which is more restrictive than the one
$Re_p/\kappa  \ll 1$ required by  Vasseur and Cox where $\kappa=2a/H$ is the confined ratio,
$H$ being the distance between two walls.
Direct numerical simulations have been used for understanding particle
motion in shear flows. Feng et al. \cite{FengJ1994} investigated the motion
of neutrally buoyant and non-neutrally buoyant circular particle in plane
shear and Poiseuille flows using a finite element method and obtained
qualitative agreement with the results of perturbation theories and of
experiments. The numerical results of a neutrally buoyant circular cylinder 
in a shear flow of $Re_p=0.625$ have been discussed in details. The cylinder migrates 
back to the midway between two walls due to the wall repulsion at the small Reynolds 
number. They have suggested that that three factors, namely the wall repulsion due to a 
lubrication effect, the slip velocity, and the Magnus type of lift, are possible responsible
for the lateral migration. 
Ding and Aidun \cite{Ding2000} studied numerically the dynamics of a cylinder of circular 
or elliptic shape suspended in shear flow at various particle Reynolds number.
They obtained the transient from being rotary to stationary as the particle Reynolds 
number is increased for an elliptic cylinder. For the cases of the circular cylinder, 
the effect of the two walls on the rotation speed $\omega$ has been studied. For the
confined ratio $\kappa=0.5$, Ding and Aidun reported $|\omega|/G \propto Re_p^{-0.28}$.
Similar result,$|\omega|/G \propto Re_p^{-0.25}$, has been observed experimentally by
Zettner and Yoda \cite{Zettner2001}.

In this paper, we first discuss the generalization of a distributed
Lagrange multiplier/fictitious domain method (DLM/FD method) developed in \cite{Pan2002} 
to non-spherical neutrally buoyant cylinders in two-dimensional shear flows and 
to the cases with zero angular velocity as a constraint.
Similar DLM/FD methods for freely moving neutrally buoyant particle has been 
successfully applied, in \cite{Pan2008, Pan2005, Yang2005}, to simulate particulate 
flow in two and three dimensions with neutrally buoyant particles. 
But for the cases of a neutrally buoyant elliptic cylinder in two dimensional flows, 
the methodology has not been fully validated yet. 
We have validated the numerical method by
comparing with the computational results in Ding and Aidun \cite{Ding2000} for a 
cylinder of circular and elliptic shape and the experimental results in 
Zettner and Yoda \cite{Zettner2001} for a circular cylinder. On the wall effect on 
the angular velocity of the circular cylinder, we have obtained
$|\omega|/G \propto Re_p^{-0.2771}$ for the confined ratio $\kappa=0.5$ 
which is in a good agreement with the results obtained by Ding and Aidun \cite{Ding2000} 
and Zettner and Yoda \cite{Zettner2001}. We have also studied the wall effect on the 
angular velocity of the elliptic cylinder 
which is more complicated due to the non-circular shape.

\begin{figure} 
\begin{center}
\leavevmode
\epsfxsize=4.5in
\epsffile{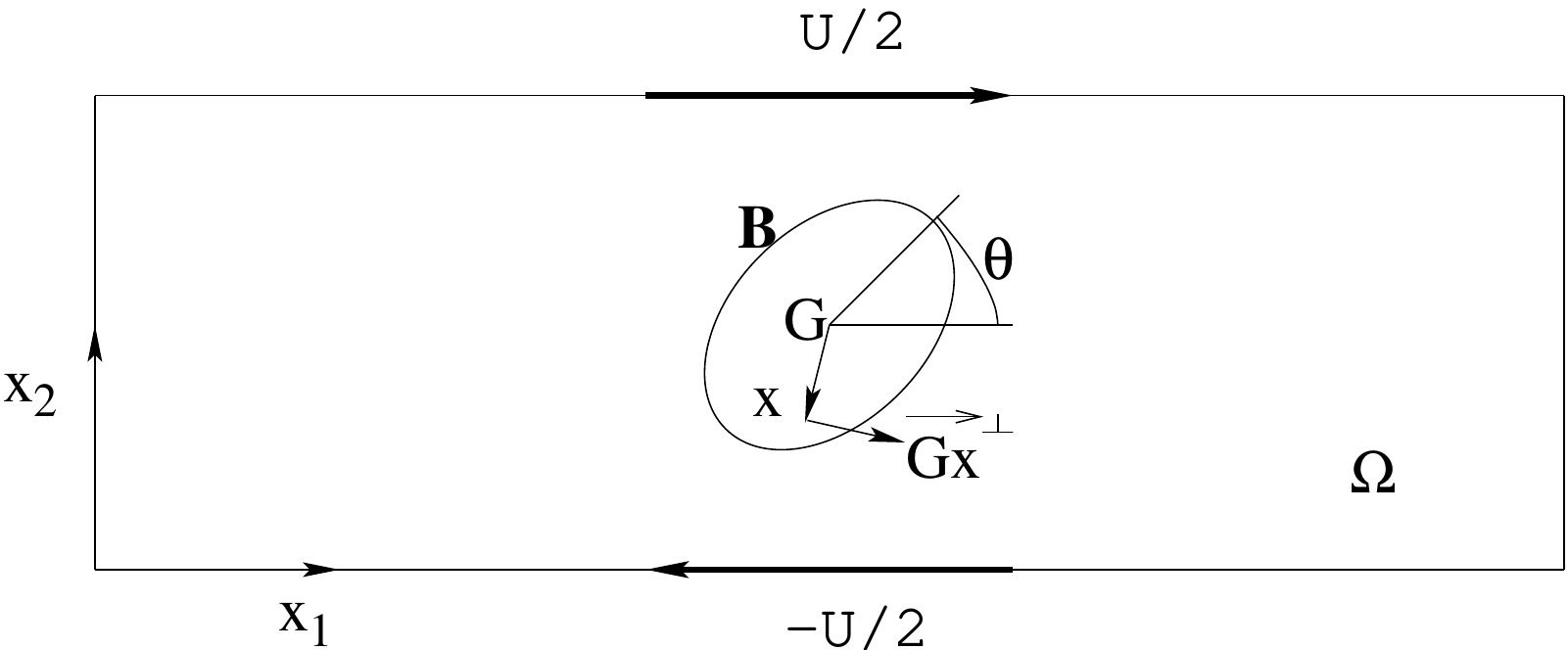}
\end{center}
\caption{ An example of two-dimensional flow region with a rigid body.}\label{fig:1}
\end{figure}

Concerning the equilibrium position of a neutrally buoyant circular cylinder in shear flow,
recent studies by Cherukat, McLaughlin and Dandy \cite{Cherukat1999} and Kurose and 
Komori \cite{Kurose1999} focus on lift and drag on a stationary sphere in unbounded linear shear flow. 
The equilibrium positions have not been studied in these works. Feng and Michaelides \cite{Feng2003} 
have investigated the equilibrium positions of non-neutrally buoyant circular cylinders in 
two-dimensional shear flow. In their simulations, the density ratio between the solid and fluid 
is between 1.005 and 1.1. The equilibrium heights of their lightest circular cylinder (the density
ratio of 1.005) are far below the centerline. We have obtained that
the cylinder stays in the middle between two walls as expected when placing 
it there initially. But when the initial position of the mass center of a circular cylinder is away 
from the centerline, the equilibrium position depends on the particle Reynolds number $Re_p$ and 
the confined ratio $\kappa$. For those placing away from the centerline initially,
the circular cylinder migrates back to the centerline for $Re_p < Re_{p,c}$ 
where $Re_{p,c}$ is the critical value. For $Re_p > Re_{p,c}$, the equilibrium position 
is between the wall and the centerline. The critical particle 
Reynolds number is increasing when increasing the confined ratio. 
Concerning the Magnus lift effect on the equilibrium position, we have added a constraint, 
zero angular velocity, to the motion of a circular cylinder and obtained that the equilibrium 
position of the circular cylinder moving with zero angular velocity is lower than those of 
freely moving cylinder when both are away from the middle. 
These results show that the Magnus lift does play 
a role as expected. Also from the computed slip velocity of the circular cylinder, it
shows that the circular cylinder lags the fluid, at least for $Re_p > Re_{p,c}$,
which means that there is a force pushing the cylinder toward the wall (see Fig. \ref{fig:1} for the 
setup of the boundary conditions). 
Hence the equilibrium position of the cylinder is up to the 
interplay between the slip velocity, the Magnus lift and the wall repulsion force.
The content of this paper is as follows: In Section 2 we introduce a
fictitious domain formulations of the model problem
associated with the neutrally buoyant long particle cases; then in Section
3 we discuss the time and space discretization and in Section 4 we
present and discuss the numerical results.

\vskip 8ex
\noindent{\large\bf 2. A fictitious domain formulation of the model problem}
\vskip 2ex

A fictitious domain formulation
with distributed Lagrange multipliers for flow around freely moving
particles and its associated computational methods have
been developed and tested in, e.g., \cite{RG1999, RG2001, Pan2001, Pan2002a}. For the cases
of neutrally buoyant particles, similar methodologies have been developed in
\cite{Pan2008, Pan2002, Pan2005, Yang2005}.
But for the cases of a neutrally buoyant elliptic cylinder in two dimensional flows, the  
methodology has not been discussed and fully validated yet. In this paper, we first discuss
the formulation for the case of a neutrally buoyant elliptic cylinder and then present
the numerical results to validate the methodology.

Let $\Omega \subset \R^2$ be a rectangular region
filled with a Newtonian viscous incompressible fluid (of
density $\rho$ and dynamic viscosity $\mu$) and containing
a freely moving neutrally buoyant rigid particle  $B$ centered at $\G=\{G_1, G_2\}^t$
of density $\rho$. The flow is modeled
by the Navier-Stokes equations and the motion of the particle $B$
is described by the Euler-Newton's equations. We define
\begin{eqnarray*}
&&W_{{\bf g}_0, p} = \{\bv|\bv \in (H^1(\Omega))^2, \ \bv = {\bf g}_0 \
\text{\it on the top and bottom of $\Omega$ and} \\
&&\hskip 55pt \bv \ \text{\it is periodic in the $x_1$ direction} \},\\
&&W_{0, p} = \{\bv|\bv \in (H^1(\Omega))^2, \ \bv = {\bf 0} \
\text{\it on the top and bottom of $\Omega$ and} \\
&&\hskip 55pt \bv \ \text{\it is periodic in the $x_1$ direction} \},\\
&&L_{0}^2  = \{q|q \in L^2(\Omega), \int_{\Omega} q\, d\x=0,\},\\
&&\Lambda_0(t) = \{\bmu| \bmu \in (H^1(B(t)))^2, <\bmu,{\bf e}_i>_{B(t)}=0, \ i=1, 2,
<\bmu, \Gx >_{B(t)} = 0\}
\end{eqnarray*}
with ${\bf e}_1=\{1, 0\}^t$, ${\bf e}_2=\{0, 1\}^t$, $\Gx =\{-(x_2-G_2),x_1-G_1\}^t$ and
$<\cdot,\cdot>_{B(t)}$ an inner product on $\Lambda_0(t)$ which can
be the standard inner product on $(H^1(B(t)))^2$.  For simple shear flow, we have
${\bf g}_0=(U/2,0)^t$ on the top wall and $(-U/2,0)^t$ on the bottom wall. Then the fictitious domain formulation
with distributed Lagrange multipliers for flow around a freely moving  neutrally buoyant
particle of the elliptic shape is as follows
\begin{eqnarray*}
&&For \ a.e. \ t>0, \ find \
\bu(t) \in W_{{\bf g}_0, p}, \ p(t) \in L_{0}^2, \ \V_{\G}(t) \in
\R^2, \ \G(t) \in \R^2, \\
&&\omega(t) \in \R, \ \theta(t) \in \R, \ \blambda(t) \in \Lambda_0(t) \ \ \text{\it such that}
\end{eqnarray*}
\begin{eqnarray}
&&\begin{cases}
\rho \into \left[\dfrac{\partial \bu}{\partial t}+(\bu \cdot \bnabla) \bu \right] \cdot \bv\ d\x
+ \mu \into \bnabla \bu : \bnabla \bv\, d\x
- \into p \bnabla \cdot \bv\, d\x \\
\ \ \ = <\blambda, \bv>_{B(t)}, \
\forall \bv \in W_{0, p},\end{cases} \label{eqn:2.1}\\
&& \into q \bnabla \cdot \bu(t) d\x = 0, \ \forall q \in L^2(\Omega),\label{eqn:2.2}\\
&&<\bmu, \bu(t) >_{B(t)} =0, \ \forall \bmu \in \Lambda_0(t),\label{eqn:2.3}\\
&&\dfrac{d\G}{dt}=\V_{\G},  \label{eqn:2.4} \\
&&\dfrac{d\theta}{dt}=\omega,\label{eqn:2.4a} \\
&& \V_{\G}(0) = \V_{\G}^0, \ \omega(0) = \omega^0, \ \G(0) = \G^0=\{G^0_1, G^0_2\}^t, 
\ \theta(0)=\theta^0,
\label{eqn:2.5}\\
&&\bu(\x, 0) = {\overline \bu}_0(\x) = \begin{cases}
\bu_0(\x), \ \forall \x \in \obo, \\
\V_{\G}^0 + \omega^0 \{-(x_2-G^0_2),x_1-G^0_1\}^t, \ \forall \x \in \overline{B(0)},
\end{cases}
\label{eqn:2.6}
\end{eqnarray}
\noindent where
$\bu$ and $p$ denote velocity and pressure, respectively,
$\blambda$ is a Lagrange multiplier,
$\V_{\G}$ is the translation velocity of the particle $B$, 
$\omega$ is the angular velocity of $B$, 
and $\theta$ is the angle between the horizontal
direction and the long axis of the elliptic cylinder (see Fig.\ref{fig:1}).
We suppose that the no-slip condition holds on $\partial B$.
We also use, if necessary, the notation $\phi(t)$ for the function $\x \to
\phi(\x,t)$.

\begin{remark}
The hydrodynamical forces and torque imposed on the rigid body
by the fluid are built in (\ref{eqn:2.1})-(\ref{eqn:2.6})
implicitly (see \cite{RG1999, RG2001} for details), thus
we do not need to compute them explicitly in the simulation.
Since in (\ref{eqn:2.1})-(\ref{eqn:2.6}) the flow field is
defined on the entire domain $\Omega$, it can be computed
with a simple structured grid.
\end{remark}
\begin{remark}
In (\ref{eqn:2.3}), the rigid body motion in the region occupied
by the particle is enforced via the Lagrange multiplier $\blambda$.
To recover the translation velocity $\V_{\G}(t)$  and the angular
velocity $\omega(t)$, we solve the following equations
\begin{eqnarray}
&&<{\bf e}_i, \bu(t)-\V_{\G}(t)- \omega(t) \ \Gx >_{B(t)}=0,\ for \ i=1, 2,\label{eqn:2.7}\\
&&<\Gx , \bu(t)-\V_{\G}(t)- \omega(t) \ \Gx >_{B(t)}=0.\label{eqn:2.8}
\end{eqnarray}
\end{remark}

\begin{remark} To investigate the effect of the Magnus type of lift on the lateral migration
of the cylinder, we have considered the cases of the cylinder freely moving in shear 
flow with zero angular velocity. For this special consideration, we have the following
modified formulation
\begin{eqnarray*}
&&For \ a.e. \ t>0, \ find \
\bu(t) \in W_{{\bf g}_0, p}, \ p(t) \in L_{0}^2, \ \V_{\G}(t) \in
\R^2, \ \G(t) \in \R^2, \\
&&\blambda(t) \in \Lambda_0(t) \ \ \text{\it such that}
\end{eqnarray*}
\begin{eqnarray}
&&\begin{cases}
\rho \into \left[\dfrac{\partial \bu}{\partial t}+(\bu \cdot \bnabla) \bu \right] \cdot \bv\ d\x
+ \mu \into \bnabla \bu : \bnabla \bv\, d\x
- \into p \bnabla \cdot \bv\, d\x \\
\ \ \ = <\blambda, \bv>_{B(t)}, \
\forall \bv \in W_{0, p},\end{cases} \label{eqn:2.9}\\
&& \into q \bnabla \cdot \bu(t) d\x = 0, \ \forall q \in L^2(\Omega),\label{eqn:2.10} \\
&&<\bmu, \bu(t) >_{B(t)} =0, \ \forall \bmu \in \Lambda_0(t),\label{eqn:2.11}\\
&&\dfrac{d\G}{dt}=\V_{\G},  \label{eqn:2.12} \\
&& \V_{\G}(0) = \V_{\G}^0,  \ \G(0) = \G^0=\{G^0_1, G^0_2\}^t,
\label{eqn:2.13}\\
&&\bu(\x, 0) = {\overline \bu}_0(\x) = \begin{cases}
\bu_0(\x), \ \forall \x \in \obo, \\
\V_{\G}^0, \ \forall \x \in \overline{B(0)}
\end{cases}
\label{eqn:2.14}
\end{eqnarray}
with the modified multiplier space
\begin{equation*}
\Lambda_0(t) = \{\bmu| \bmu \in (H^1(B(t)))^2, <\bmu,{\bf e}_i>_{B(t)}=0, \ i=1, 2\}.
\end{equation*}
The translation velocity $\V_{\G}(t)$ is recovered via
\begin{equation}
<{\bf e}_i, \bu(t)-\V_{\G}(t)>_{B(t)}=0,\ for \ i=1, 2.\label{eqn:2.15}
\end{equation}
\end{remark}

\vskip 4ex
\noindent{\large\bf 3. Space approximation and time discretization}
\vskip 2ex
Concerning the {\it space approximation} of problem (\ref{eqn:2.1})-(\ref{eqn:2.6}) by a
finite element method, we have used $P_1$-$iso$-$P_2$ and $P_1$ finite elements for the
velocity field and pressure, respectively (like in Bristeau et al. \cite{Bristeau1987}).
We approximate then $W_{{\bf g}_0, p}$,  $W_{0, p}$, $L^2$ and
$L^2_{0}$ by the following finite dimensional spaces
\begin{eqnarray}
W_{{\bf g}_0,h}=\{\bv_h &|& \bv_h \in (C^0(\overline{\Omega}))^2, \
\bv_h|_T \in P_1\times P_1, \ \forall T\in \ct_h,  \ \bv_h = {\bf g}_0 \
\text{\it on the top}\label{eqn:3.1a}\\
&&\text{\it and bottom of $\Omega$ and} \ \bv \ \text{\it is periodic at $\Gamma$ in the $x_1$
direction} \ \}, \nonumber \\
W_{0,h}=\{\bv_h &|& \bv_h \in (C^0(\overline{\Omega}))^2, \
\bv_h|_T \in P_1\times P_1, \ \forall T\in \ct_h,  \ \bv_h = {\bf 0} \
\text{\it on the top}\label{eqn:3.1}\\
&&\text{\it and bottom of $\Omega$ and} \ \bv \ \text{\it is periodic at $\Gamma$ in the $x_1$
direction} \ \}, \nonumber
\end{eqnarray}
\begin{eqnarray}
L^2_{h}=\{q_h &|& q_h\in C^0(\overline{\Omega}), \ q_h|_T\in P_1,
\ \forall T\in \ct_{2h}, \ q_h \ \text{\it is periodic}\label{eqn:3.2} \\
&&\text{\it at $\Gamma$ in the $x_1$ direction}\}, \nonumber
\end{eqnarray}
and
\begin{equation}
L^2_{0,h}=\{q_h|q_h\in L^2_{h}, \ \int_{\Omega} q_h \, d\x=0 \},
\label{eqn:3.3}
\end{equation}
respectively; in (\ref{eqn:3.1a})-(\ref{eqn:3.3}), $P_1$ is the space of
polynomials in two variables of degree $\le 1$.

A finite dimensional space approximating $\Lambda_0(t)$
is defined as follows:  let $\{\x_i\}_{i=1}^{N}$ be a set of points
covering $\overline{B(t)}$; the discrete multiplier space $\Lambda_{h}(t)$ 
is defined by
\begin{equation}
\Lambda_{h}(t)=\{\bmu_h|\bmu_h=\dsum_{i=1}^{N}\bmu_i
\delta(\x-\x_i), \ \bmu_i\in\R^2, \ \forall i=1,...,N\},
\label{eqn:3.4}
\end{equation}
where $\delta(\cdot)$ is the Dirac measure at $\x={\bf 0}$.
Then, we have the inner product defined by 
\begin{equation}
<\bmu_h,\bv_h>_{B_h(t)} = \dsum_{i=1}^{N}
\bmu_i\cdot\bv_h(\x_i), \ \forall \bmu_h \in \Lambda_{h}(t),
\ \bv_h\in W_{0,h}
\label{eqn:3.5}
\end{equation}
and approximate $\Lambda_0(t)$ by
\begin{equation}
\Lambda_{0,h}(t) =\{\mu_h| \mu_h \in \Lambda_h(t), \ <\mu_h,{\bf e}_i>_{B_h(t)}=0,\
i=1, 2, \ <\mu_h, \Gx >_{B_h(t)}=0 \}.
\end{equation}

Using the above finite dimensional spaces leads to the
following approximation of problem   (\ref{eqn:2.1})-(\ref{eqn:2.6}):
\begin{eqnarray*}
&&For \ a.e. \ t>0, \ find \
\bu(t) \in W_{{\bf g}_0,h}, \ p(t) \in L_{0,h}^2, \ \V_{\G}(t) \in
\R^2, \ \G(t) \in \R^2, \\
&&\omega(t) \in \R, \ \theta(t) \in \R,\ \blambda_h(t) \in \Lambda_{0,h}(t) \ \ \text{\it such that}
\end{eqnarray*}
\begin{eqnarray}
&&\begin{cases}
\rho \into \left[\dfrac{\partial \bu_h}{\partial t}  +
  (\bu_h \cdot \bnabla) \bu_h \right] \cdot \bv\, d\x
+ \mu \into \bnabla \bu_h \boldsymbol{:} \bnabla \bv\, d\x \\
\hskip 20pt - \into p_h \bnabla \cdot \bv\, d\x = <\blambda_h, \bv>_{B_h(t)}, \
\forall \bv \in W_{0,h},\end{cases} \label{eqn:3.6}\\
&& \into q \bnabla \cdot \bu_h(t) d\x = 0, \ \forall q \in L^2_{h},\label{eqn:3.7} \\
&&<\bmu, \bu_h(t) >_{B_h(t)} =0, \ \forall \bmu \in \Lambda_{0,h}(t),\label{eqn:3.8}\\
&&\dfrac{d\G}{dt}=\V_{\G},  \label{eqn:3.9} \\
&&\dfrac{d\theta}{dt}=\omega,\label{eqn:3.9a} \\
&& \V_{\G}(0) = \V_{\G}^0, \ \omega(0) = \omega^0, \ \G(0) = \G^0=\{G^0_1, G^0_2\}^t, 
\ \theta(0)=\theta^0, \label{eqn:3.10}\\
&&\bu_h(\x, 0) ={\overline\bu}_{0,h}(\x) \label{eqn:3.11} \
(\text{with} \bnabla \cdot {\overline\bu}_{0,h} =0).
\end{eqnarray}

Applying a first order operator splitting scheme \`a la
Marchuk-Yanenko \cite{Marchuk1990}  to the equations
(\ref{eqn:3.6})-(\ref{eqn:3.11}) at each time step and the Euler backward method in time
for some subproblems, we obtain (after
dropping some of the subscripts $h$):
\vskip 1ex
\begin{equation}
\bu^0={\overline\bu}_{0,h}, \ \V_{\G}^0, \ \omega^0, \ \G^0, \ and \ \theta^0 \ are \ given;
\label{eqn:3.12}
\end{equation}
\vskip 1ex
\noindent{\it For $n \ge 0$, knowing $\bu^n$, $\V_{\G}^n$, $\omega^n$, $\G^n$, and $\theta^n$,
compute  $\bu^{n+1/6}$ and $p^{n+1/6}$ via the solution of}
\begin{equation}
\begin{cases}
\displaystyle \rho \into
\frac{\bu^{n+1/6} - \bu^n}{\triangle t} \cdot \bv\, d\x
- \into p^{n+1/6} \bnabla \cdot \bv\, d\x=0,
\  \forall \bv \in W_{0,h}, \\
\displaystyle \into q \bnabla \cdot \bu^{n+1/6}\,d\x=0,
\ \forall q \in L^2_{h}; \ \bu^{n+1/6} \in W_{{\bf g}_0,h},
\  p^{n+1/6} \in L^2_{0,h}.
\end{cases}  \label{eqn:3.13}
\end{equation}
\vskip 4ex
\noindent {\it Then compute $\bu^{n+2/6}$ via the solution of}
\vskip 2ex
\begin{eqnarray}
&&\begin{cases}
\displaystyle  \into \frac{\partial \bu}{\partial t} \cdot \bv\, d\x
 + \into ( \bu^{n +1/6} \cdot \bnabla ) \bu \cdot \bv\, d\x =0, \
\forall \bv \in W_{0,h}, \ on \ (t^n, t^{n+1}), \\
\displaystyle  \ \bu(t^n)= \bu^{n+1/6}; \ \ \bu(t) \in W_{{\bf g}_0,h},
\end{cases} \label{eqn:3.14} \\
&&\bu^{n+2/6} = \bu(t^{n+1}). \label{eqn:3.15}
\end{eqnarray}
\vskip 4ex
\noindent {\it Next, compute $\bu^{n+3/6}$ via the solution of}
\begin{equation}
\begin{cases}
\displaystyle\rho \into\frac{\bu^{n+3/6} - \bu^{n+2/6}}{\triangle t} \cdot \bv\, d\x
+\alpha \mu\into \bnabla\bu^{n+3/6} \cdot \bnabla\bv  \,d\x
=0,  \\
\forall \bv \in W_{0,h}; \  \ \bu^{n+3/6} \in W_{{\bf g}_0,h}.
\end{cases}  \label{eqn:3.16}
\end{equation}
\vskip 1ex
\noindent {\it Now predict the position and the orientation of the particle
via:}
\begin{eqnarray}
&&\dfrac{d\G}{dt}=\V_{\G}^n/2, \ on \ (t^n, t^{n+1}) \label{eqn:3.17} \\
&&\dfrac{d\theta}{dt}=\omega^n/2,\ on \ (t^n, t^{n+1})\label{eqn:3.17a} \\
&&\G(t^n) = \G^n, \ \theta(t^n) = \theta^n. \label{eqn:3.18}
\end{eqnarray}
\vskip 1ex
\noindent {\it Then set ${\G}^{n+4/6}={\G}(t^{n+1})$ and  ${\theta}^{n+4/6}={\theta}(t^{n+1})$.}
\vskip 1ex
\noindent {\it Now, compute $\bu^{n+5/6}$, ${\blambda}^{n+5/6}$,
$\V_{\G}^{n+5/6}$,  and $\omega^{n+5/6}$ via the solution of}
\vskip 1ex
\begin{equation}
\begin{cases}
\displaystyle\rho  \into   \frac{\bu^{n+5/6}
- \bu^{n+3/6}}{\triangle t}  \cdot \bv\, d\x
+\beta \mu\into \bnabla\bu^{n+5/6} \cdot \bnabla\bv\,d\x  \\
\ =<\blambda, \bv >_{B_h^{n+4/6}},
\ \forall \bv \in W_{0,h},  \\
 <\bmu, \bu^{n+5/6} >_{B_h^{n+4/6}}=0,
\ \forall {\bmu} \in \Lambda_{0,h}^{n+4/6};
\displaystyle  \ \bu^{n+5/6} \in W_{{\bf g}_0,h},
{\blambda}^{n+5/6} \in  \Lambda_{0, h}^{n+4/6},
\end{cases}  \label{eqn:3.22}
\end{equation}
\vskip 1ex
\noindent{\it and solve for $\V_{\G}^{n+5/6}$ and $\omega^{n+5/6}$ from}
\begin{equation}
\begin{cases}
<{\bf e}_i, \bu^{n+5/6} -\V_{\G}^{n+5/6} -
\omega^{n+5/6} \ {\overrightarrow{\G^{n+4/6}\x}}^{\perp} >_{B_h^{n+4/6}}=0,
\ for \ i=1, 2,\\
<{\overrightarrow{\G^{n+4/6}\x}}^{\perp} , \bu^{n+5/6} -\V_{\G}^{n+5/6} -
\omega^{n+5/6} \ {\overrightarrow{\G^{n+4/6}\x}}^{\perp} >_{B_h^{n+4/6}}=0,
\end{cases}
\label{eqn:3.23}
\end{equation}
\vskip 1ex
\noindent {\it Finally, correct the position and the orientation of the particle
via:}
\begin{eqnarray}
&&\dfrac{d\G}{dt}=\V_{\G}^{n+5/6}/2, \ on \ (t^n, t^{n+1}) \label{eqn:3.25} \\
&&\dfrac{d\theta}{dt}=\omega^{n+5/6}/2,\ on \ (t^n, t^{n+1})\label{eqn:3.26} \\
&&\G(t^n) = \G^{n+4/6}, \ \theta(t^n) = \theta^{n+4/6}, \label{eqn:3.27}
\end{eqnarray}
\vskip 1ex
\noindent {\it and set ${\G}^{n+1}={\G}(t^{n+1})$ and  ${\theta}^{n+1}={\theta}(t^{n+1})$. 
Finally, we set $\bu^{n+1}=\bu^{n+5/6}$, $\V_{\G}^{n+1}=\V_{\G}^{n+5/6}$, and
$\omega^{n+1}=\omega^{n+5/6}$.}
\vskip 2ex
In above algorithm (\ref{eqn:3.12})-(\ref{eqn:3.27}),  we have
$t^{n+s}=(n+s)\triangle t$,  $\Lambda_{0,h}^{n+s}=\Lambda_{0,h}(t^{n+s})$,
$B_h^{n+s}$ is the region occupied by the particle centered at $\G^{n+s}$.
Finally, $\alpha$ and $\beta$ verify $\alpha+\beta=1$; we have chosen
$\alpha=1$ and $\beta=0$ in the numerical simulations discussed later.

At each time step we have a sequence  of subproblems (\ref{eqn:3.13}), (\ref{eqn:3.14}),
(\ref{eqn:3.16}) and (\ref{eqn:3.22}).
The degenerated quasi-Stokes problem (\ref{eqn:3.13}) is solved by a
preconditioned conjugate gradient method introduced in \cite{RG1998},
in which discrete elliptic problems from the preconditioning are solved by
a  matrix-free fast solver from FISHPAK by Adams et al. in \cite{Adam1980}.
The advection problem (\ref{eqn:3.14}) for the velocity field  is solved by
a wave-like equation method as in \cite{RG1997, Pan2000}.
Problem (\ref{eqn:3.16}) is a classical discrete elliptic problem which can be
solved by the same matrix-free fast solver.
To enforce the rigid body motion inside the region occupied by the particles,
we have applied the conjugate gradient method discussed in, e.g.,  \cite{Pan2008, Pan2002, Pan2005}.

\begin{figure}[t]
\begin{center}
\leavevmode
\epsfxsize=4.25in
\epsffile{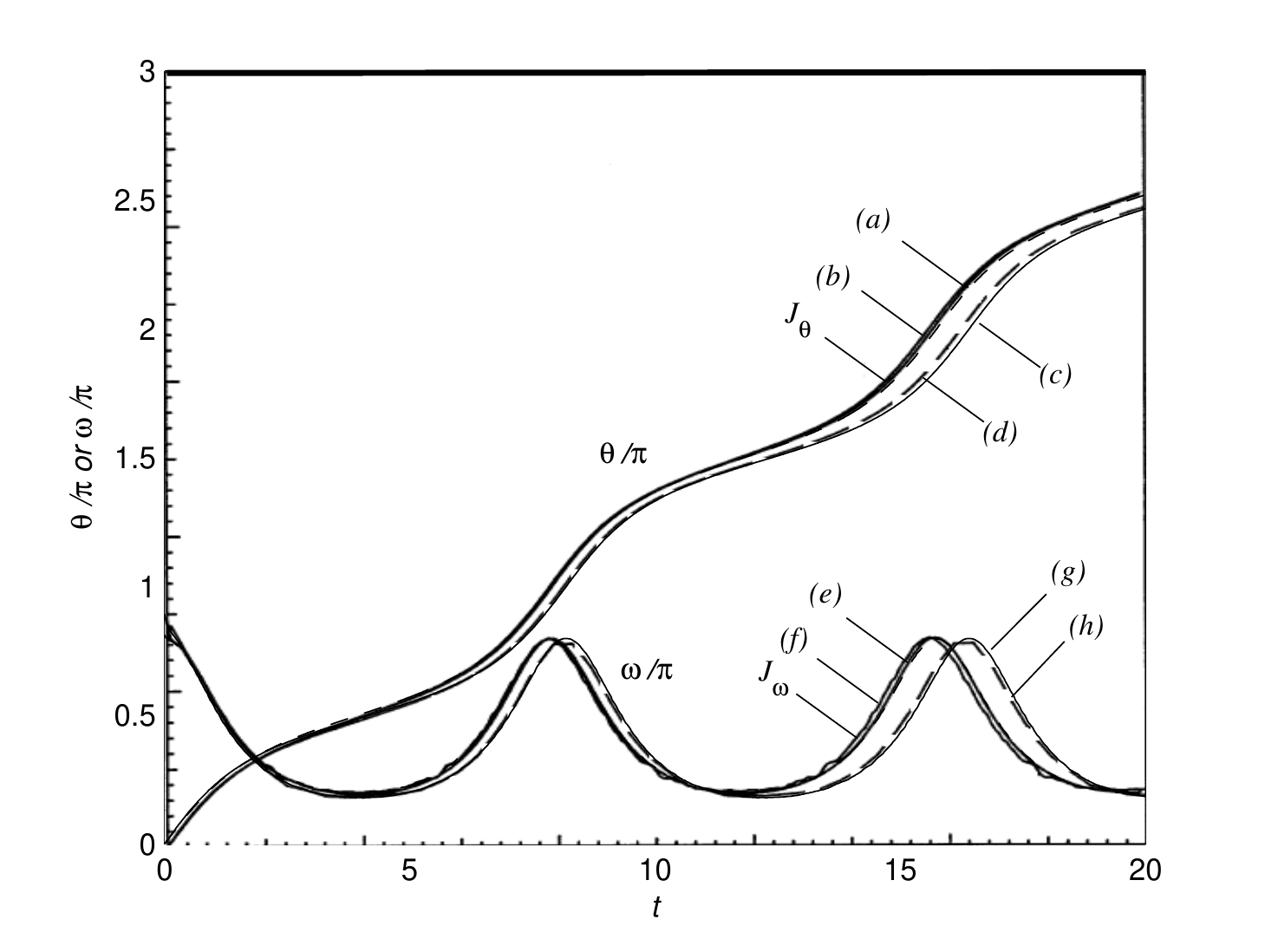}
\end{center}
\vskip -2ex
\caption{Validation of the computational results by comparing both angle of inclination and angular velocity of an elliptical cylinder. (i) The angle $\theta/\pi$: 
(a) our result (solid line) at $Re_p =$ 0.02, (b) Ding's result (dash line) at 
$Re_p =$ 0.02, (c) our result at $Re_p =$ 0.25, (d) Ding's result at $Re_p =$ 0.25. 
(ii) The angular velocity $\omega/\pi$: (e) our result at $Re_p =$ 0.02, 
(f) Ding's result at $Re_p =$ 0.02, (g) our result at $Re_p =$ 0.25, (h) Ding's result at 
$Re_p =$ 0.25.  Jeffery's solutions at $Re_p =$ 0, $J_{\theta}$ and $J_{\omega}$, are 
also plotted for comparisons.}\label{fig:2}
\end{figure}

\begin{remark} 
To solve the problem (\ref{eqn:2.9})-(\ref{eqn:2.15}) for the cases with zero
angular velocity, we have an analogue algorithm by dropping $\theta$ and $\omega$ from the algorithm (\ref{eqn:3.12})-(\ref{eqn:3.27}) and replacing 
the equation (\ref{eqn:3.23})  by
\begin{equation*}
 <{\bf e}_i, \bu^{n+5/6} -\V_{\G}^{n+5/6} >_{B_h^{n+4/6}}=0,
\ for \ i=1, 2
\end{equation*}
with the discrete multiplier space defined by
\begin{equation}
\Lambda_{0,h}(t) =\{\mu_h| \mu_h \in \Lambda_h(t), \ <\mu_h,{\bf e}_i>_{B_h(t)}=0,\
i=1, 2\}.
\end{equation}
\end{remark}

\begin{figure}[t]
\begin{center}
\leavevmode
\epsfxsize=4.25in
\epsffile{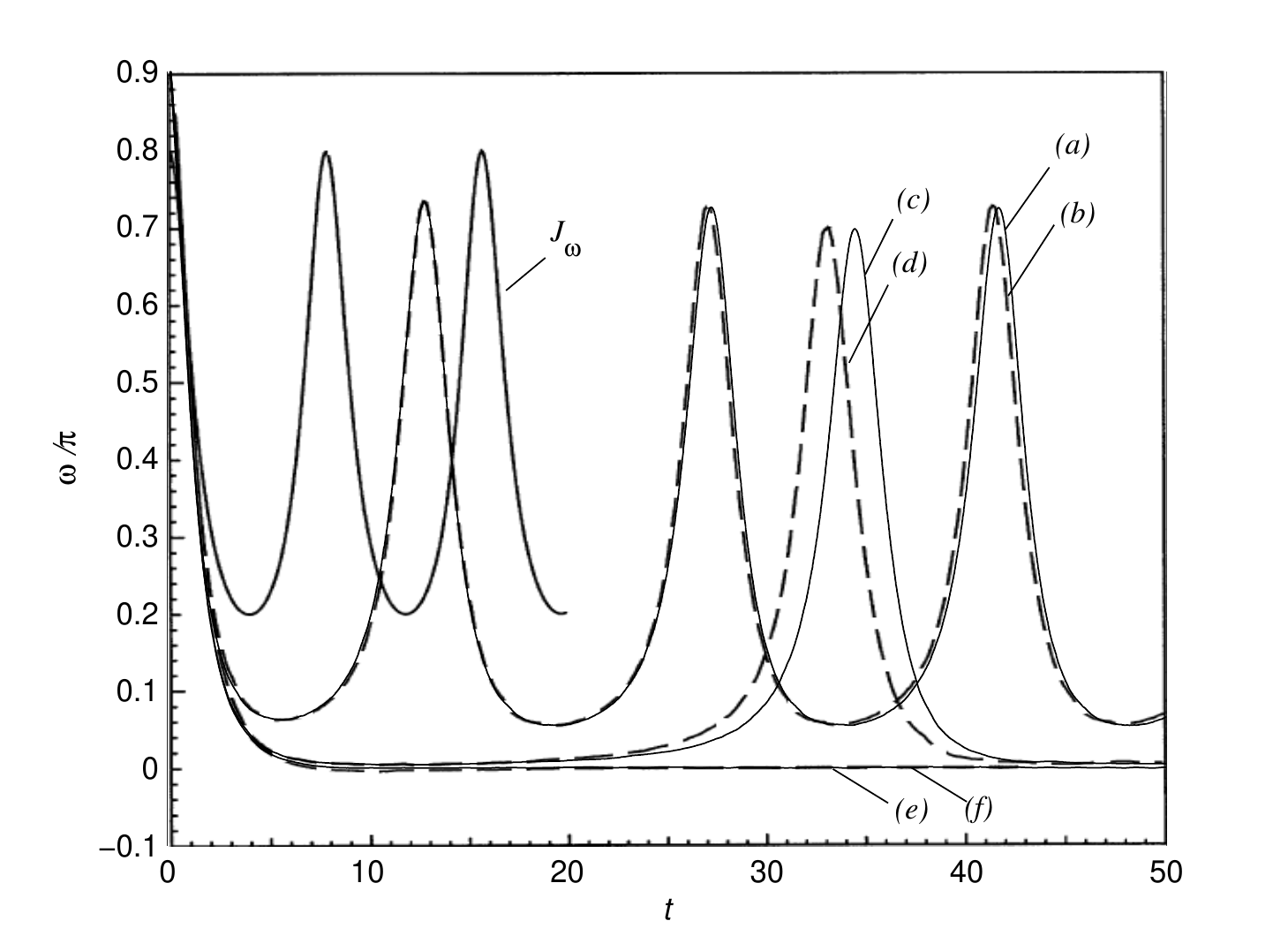}
\end{center}
\vskip -2ex
\caption{Angular velocity $\omega/\pi$ of an elliptical cylinder in shear 
flow at different $Re_p$. (a) Our result (solid line) at $Re_p =$ 3.75; (b) Ding's result 
(dash line) at $Re_p =$ 3.75; (c) Our result at $Re_p =$ 7; (d) Ding's result at $Re_p =$ 7; 
(e) Our result at $Re_p =$ 7.5 and (f) Ding's result at $Re_p =$ 7.5. 
Jeffery's solution $J_{\omega}$ at $Re_p =$ 0 is also plotted for comparisons.}\label{fig:3}
\end{figure}

\begin{figure}
\begin{center}
\leavevmode
\epsfxsize=6.in
\epsffile{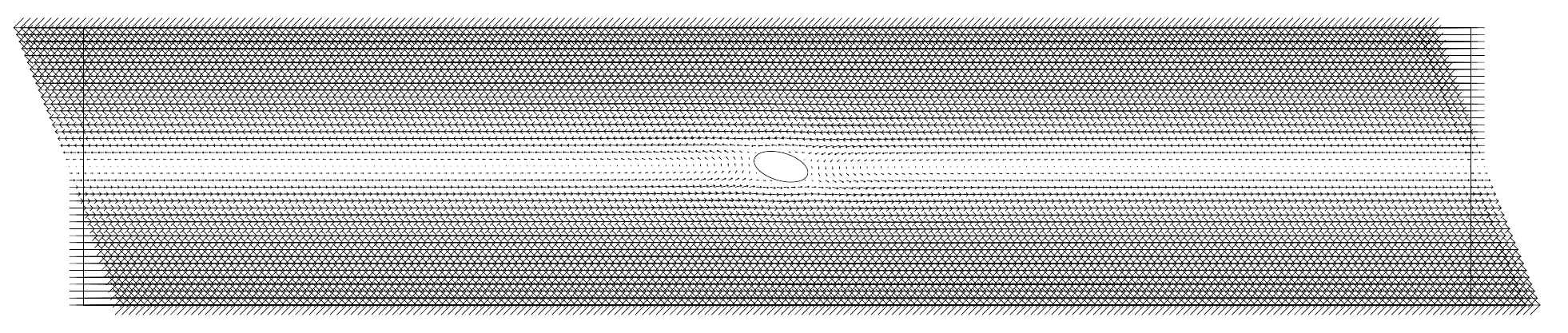}
\end{center}
\caption{The velocity field of the case of $\kappa$=0.2 and $Re_p=8.25$.} \label{fig:4}
\begin{center}
\leavevmode
\epsfxsize=2.95in
\epsffile{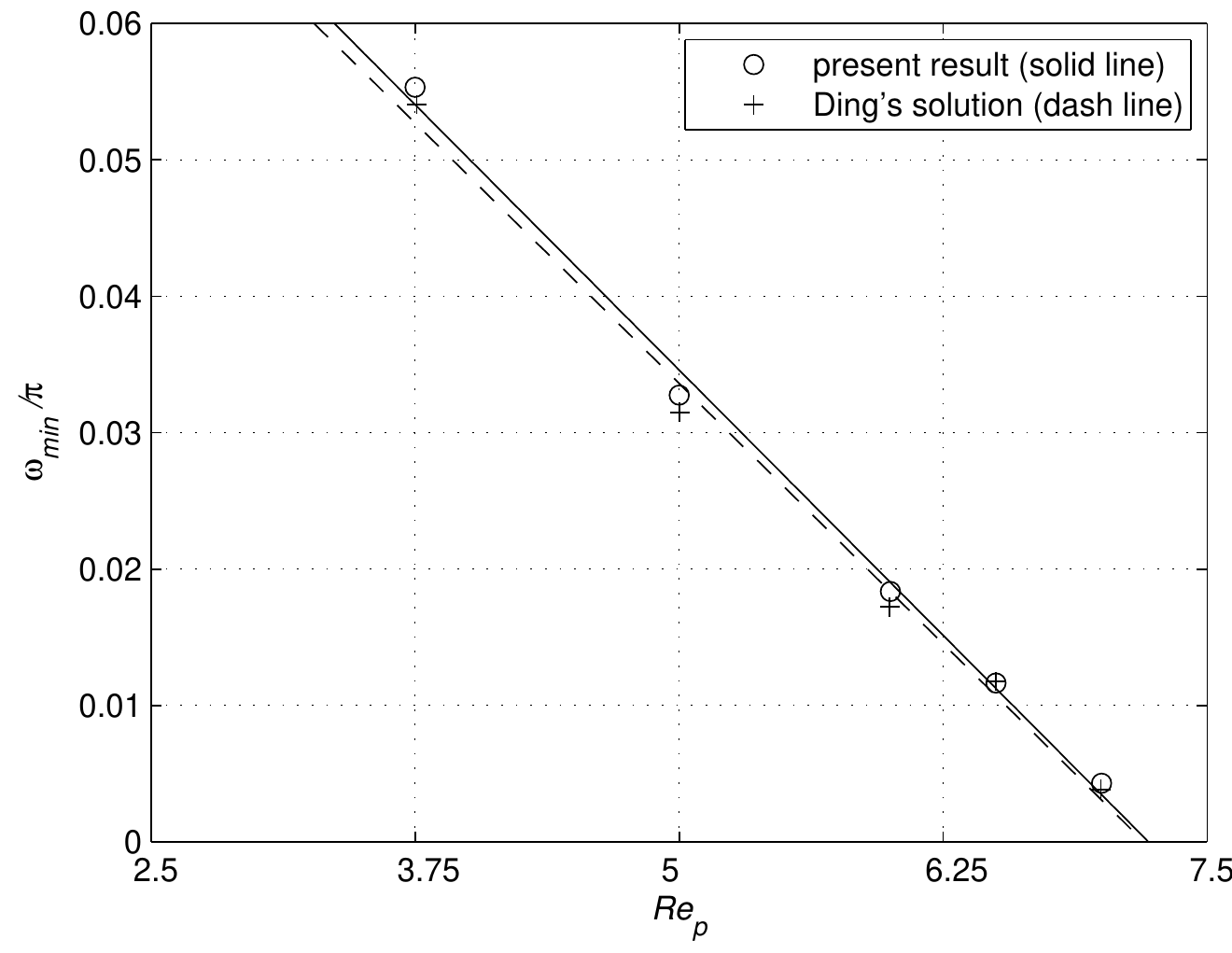}
\epsfxsize=2.95in
\epsffile{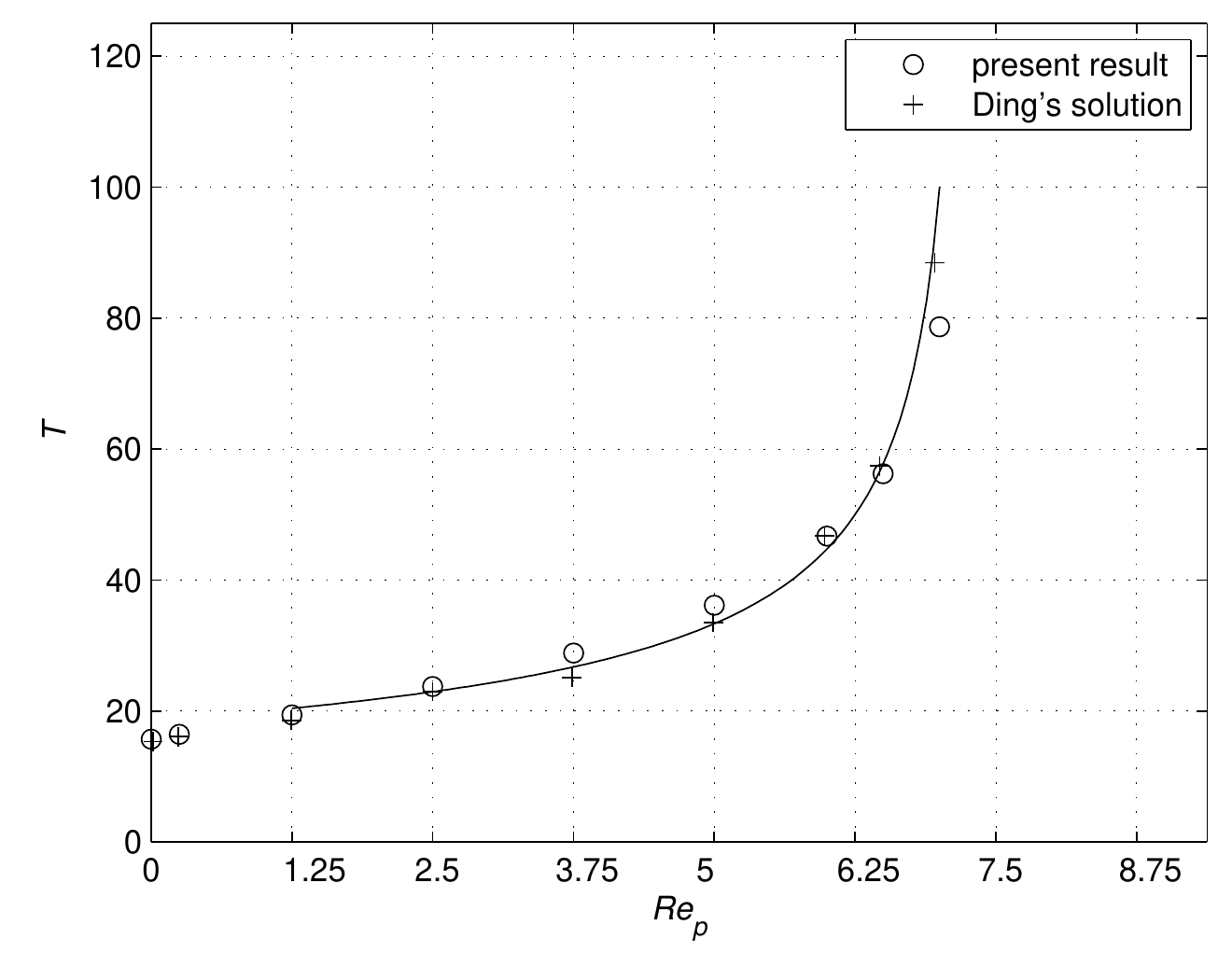}
\end{center}
\caption{Minimum angular velocity $\omega/\pi$ versus $Re_p$ of an elliptical cylinder (left).
Period of motion $T$ of an elliptical cylinder. It is noted that $T$ increases 
to infinity as $Re_p$ approaches the critical value $Re_{p,c} \sim 7.25$ (right).}\label{fig:5}
\begin{center}
\leavevmode
\epsfxsize=6.in
\epsffile{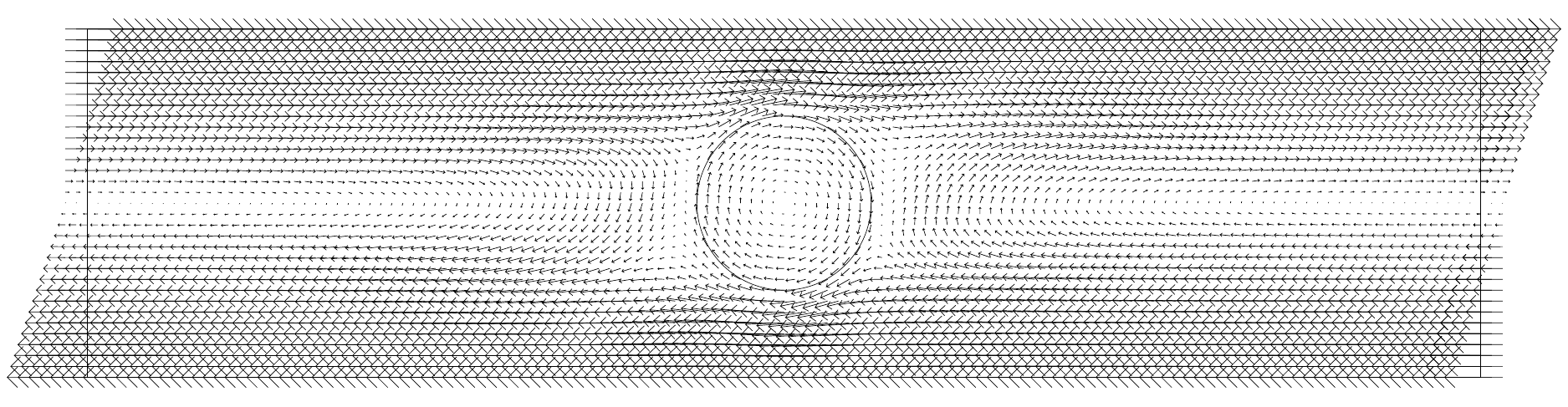}
\end{center}
\caption{The velocity field of the case $\kappa$=0.5 and $Re_p=15$.} \label{fig:6}
\end{figure}

\vskip 6ex
\noindent{\large\bf 4. Results and discussions}
\vskip 2ex
\noindent{\bf 4.1. The motion of an elliptic cylinder in linear shear flow}
\vskip 2ex

We have first considered the motion of a neutrally buoyant elliptical 
cylinder in linear shear flow  studied by Ding and Aidun \cite{Ding2000} for the 
validation purpose.  Their results were obtained by the lattice Boltzmann equation. 
The domain of computation  is $\Omega = [0,5]\times[0,1]$, then the height is $H=1$. 
The confined ratio is $\kappa= 2 a/ H = 0.2$, and  the aspect ratio is $AR=b/a = 0.5$ 
where $a$ is the length of the semi-major axis and $b$ is the length of the semi-minor axis. 
The density of the fluid is $\rho=1$ and the kinetic viscosity is determined by 
the specified value of the Reynolds number via $\nu=G a^2/{Re_p}$. The shear rate is fixed 
at $G=1$ and  $U=-G H=-1$ (so the moving directions of the two walls are the same as those in 
Ding and Aidun \cite{Ding2000}). The mesh size is $h=1/320$ and the time step 
is $\triangle t=0.001$.

In Fig. \ref{fig:2}, the computational results of the angle and angular velocity
of the elliptic cylinder are in a good agreement with the Jeffery's solution \cite{Jeffery1992} 
for $Re_p = 0$ and the ones obtained by Ding and Aidun for $Re_p = 0.02$ and 0.25, respectively. 
Fig. \ref{fig:3} shows that at various particle Reynolds numbers from 0 to 7.5, our results 
of angular velocity match very well with the ones obtained by Ding and Aidun \cite{Ding2000}. 
The motion of the ellipse for 
$Re_p \le 7$ is a periodic rotation with non-uniform angular velocity, while for 
$Re_p \ge 7.5$, the ellipse does not rotate at all; instead it takes a stationary 
orientation in shear flow (see, e.g., Ding and Aidun \cite{Ding2000} for further details).
An example of the velocity field of $Re_p=8.25$ around an elliptic cylinder
with a stable orientation is shown in Fig. \ref{fig:4}.

The minimum angular velocity decreases as the particle Reynolds number is increased with 
a nearly straight line relationship as shown in Fig. \ref{fig:5}, where $Re_{p,c}$ = 7.25 
is the critical particle Reynolds number above which the rotational motion is stopped. 
Fig. \ref{fig:5} shows that the period of rotation increases rapidly as the particle 
Reynolds number is increased close to $Re_{p,c}$, and the results are in good agreements 
with those obtained by Ding and Aidun \cite{Ding2000}. The further studies of the motion of
a cylinder of axisymmetric shape in simple shear flow will be reported in a forthcoming  paper.

\begin{figure}[t]
\begin{center}
\leavevmode
\epsfxsize=4.25in
\epsffile{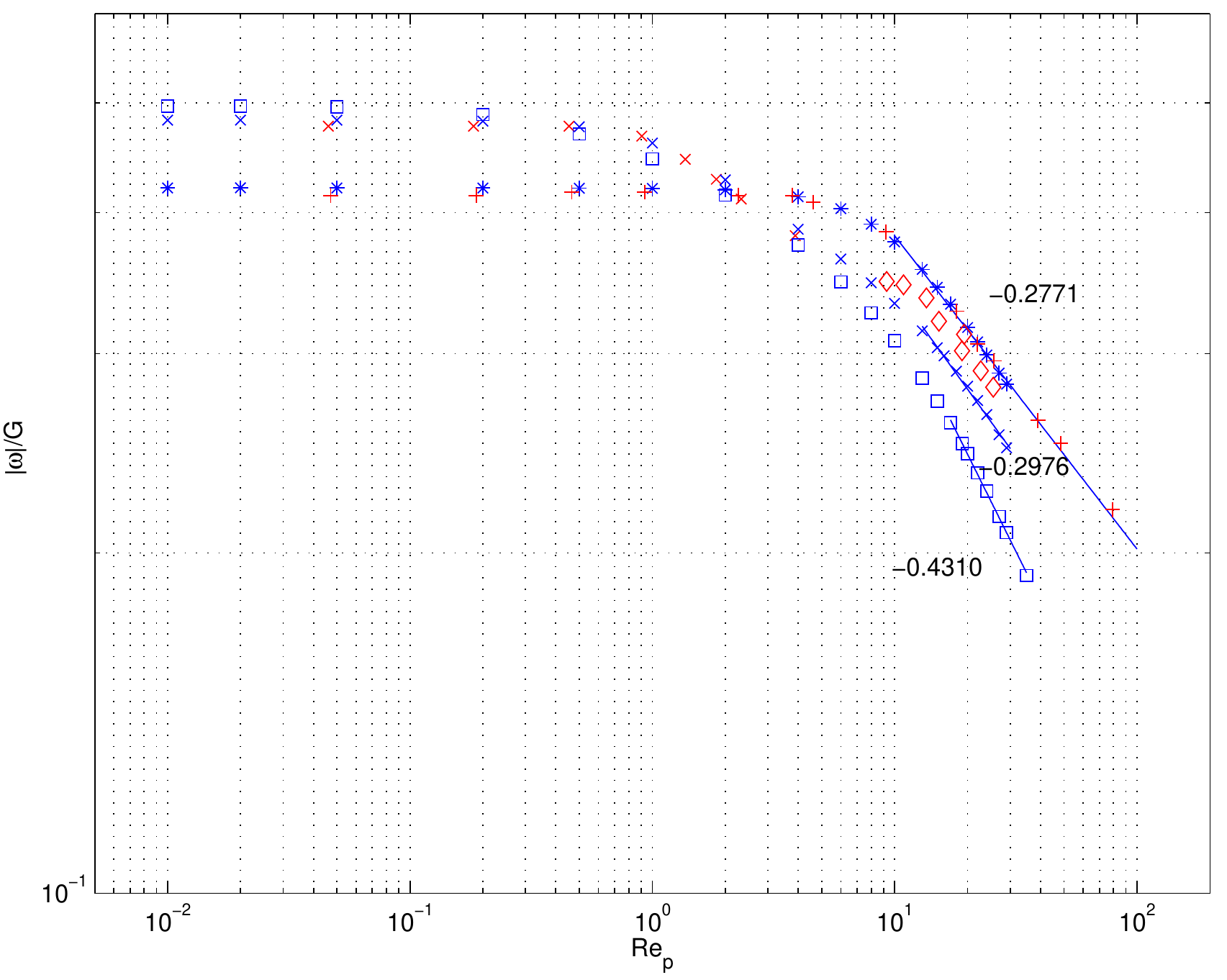}
\end{center}
\caption{Log-log plot of normalized angular velocity $\omega/G$ vs $Re_p$:
(a). $\kappa$=0.5 (blue $\ast$), 0.25 (blue $\times$), 0.125 (blue \hskip -10pt$\qed$). 
(b). Ding \& Aidun's results (2000): $\kappa=0.5$ (red $+$) and  0.25 (red $\times$). 
(c). Zettner \& Yoda's results (2001): $\kappa=0.5$ (red $\diamondsuit$).} \label{fig:7}
\end{figure}

\begin{figure}
\begin{center}
\leavevmode
\epsfxsize=2.9in
\epsffile{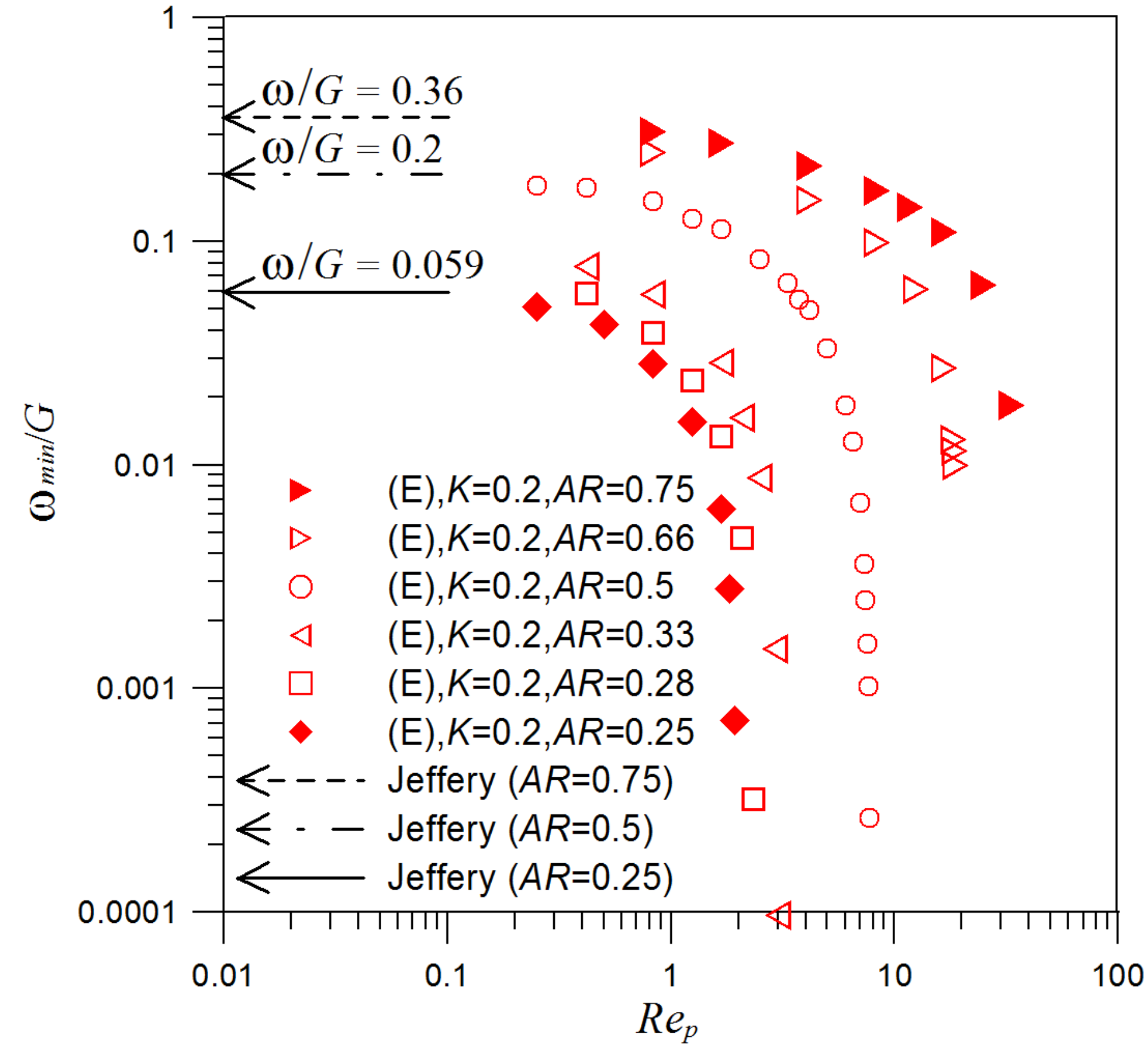} 
\epsfxsize=2.9in
\epsffile{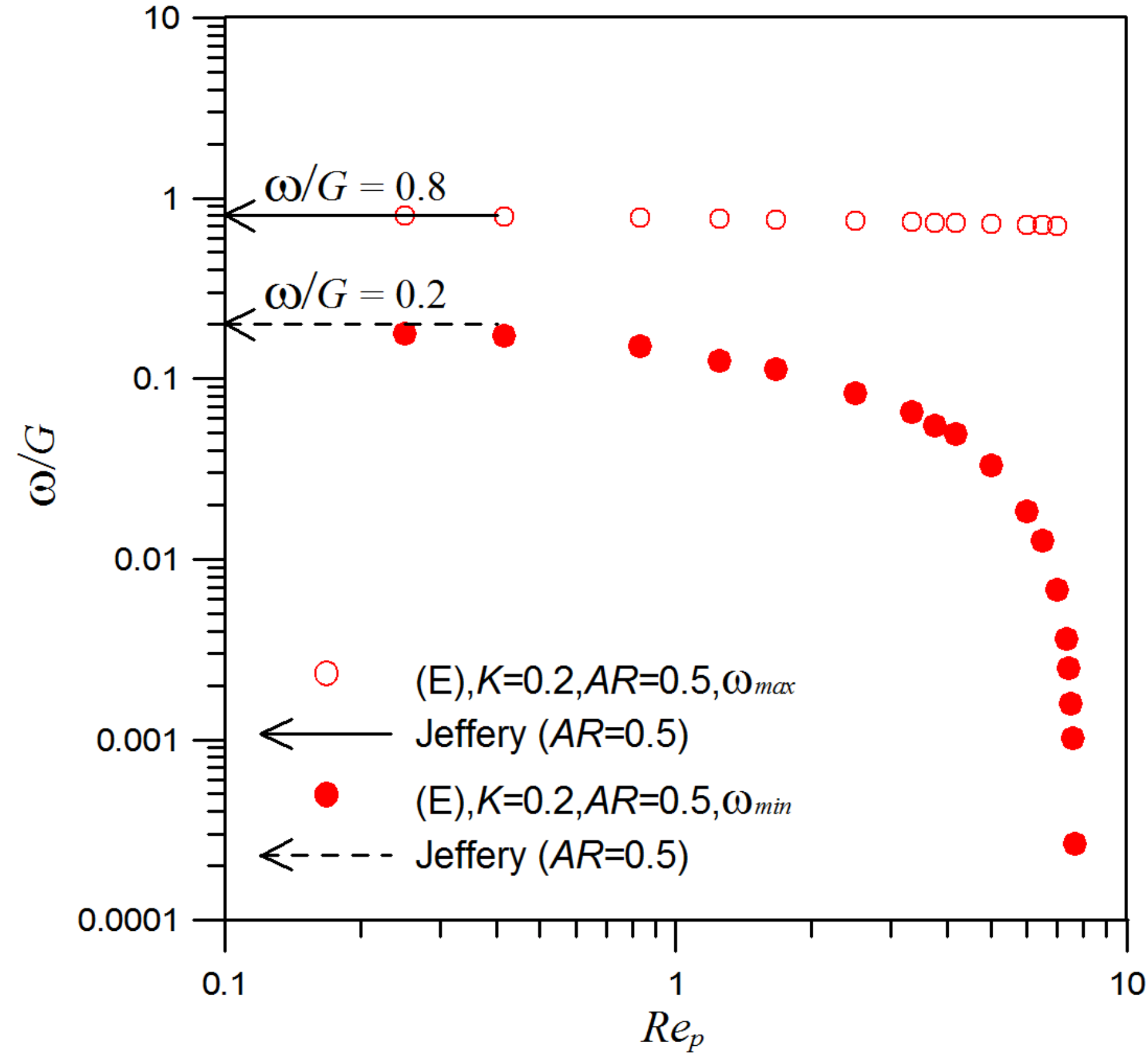}
\end{center}
\caption{Log-log plot of normalized angular velocity $|\omega_{min}|/G$ versus $Re_p$
for a fixed confined ratio $\kappa=0.2$ (left) and a case of a fixed aspect ratio $AR=0.5$ (right).} \label{fig:8}
\end{figure}

\vskip 4ex
\noindent{\bf 4.2. The rotation of a cylinder in a simple shear flow}
\vskip 2ex

In, e.g., Aidun and Ding \cite{Ding2000}, the angular velocity of a circular cylinder suspended 
in a simple shear flow has been studied via the direct numerical simulation. Experimental 
results have been also obtained in, e.g., Zettner and Yoda \cite{Zettner2001}.
In this section, we focus on the wall effect of the rotation speed of a cylinder 
freely suspended in a simple shear flow with various confined ratios. We have first 
considered the cases of a circular cylinder suspended in the middle between two walls initially. 
The domain of computation  is $\Omega = [0,L]\times[0,1]$ with $L=16a$ where $a$ is 
the circular cylinder radius. The density of the fluid is $\rho=1$ and the kinetic viscosity 
is $\nu=0.012$. The confined ratios are $\kappa=2a/H=2a=$0.125, 0.25 and 0.5. We have varied
the values of the shear rate $G$ to have different values of the particle Reynolds number $Re_p$. 
The initial position of the cylinder is at the midway between two walls.
The mass center of the circular cylinder stays at the centerline between two walls in the simulations
without giving any extra conditions for keeping it there.
A typical velocity field (here, $Re_p=15$ and $\kappa$=0.5) is shown in Fig. \ref{fig:6}.
At the zero Reynolds number, the rotation speed of a circular cylinder is $G/2$
from the Jeffery's solution.
At the small particle Reynolds numbers, the ratio $|\omega|/G$  is about to converge 
to 0.5 when decreasing the confined ratio $\kappa$ from 0.5 to 0.125 as shown 
in Fig. \ref{fig:7}. When $\kappa=$0.5, the normalized angular speed  $|\omega|/G$ is 
found to be about 0.420 for $ Re_p \le 2$. When $\kappa=$0.25, $|\omega|/G$ is about 
0.482 for $ Re_p \le 0.2$. Both are very close to those obtained in Ding and Aidun \cite{Ding2000}. 
When $\kappa=$0.125, $|\omega|/G=0.496$ is much closer to 0.5 for $ Re_p \le 0.05$ due to weaker 
effect from the walls. 

When increasing the particle Reynolds numbers, the log-log plot of $|\omega|/G$ versus $Re_p$ 
in Fig. \ref{fig:7} shows that $|\omega|/G \propto Re_p^{-0.2771}$ for $10 \le Re_p \le 29$
when $\kappa=$0.5. The one obtained in Ding and Aidun \cite{Ding2000}  is  
$|\omega|/G \propto Re_p^{-0.28}$  and the experimental results 
reported in Zettner and Yoda  \cite{Zettner2001} is $|\omega|/G \propto Re_p^{-0.25}$ for the same 
confined ratio $\kappa=$0.5. 
For the smaller values of the confined ratios, $\kappa=$0.25 and 0.125, we have obtained 
$|\omega|/G \propto Re_p^{-0.2976}$ and $|\omega|/G \propto Re_p^{-0.4310}$ respectively. 
The results of  $\kappa=$0.5 and 0.25 are in a good agreement with those obtained in  
Ding and Aidun \cite{Ding2000}.

For an elliptic cylinder suspended in linear shear flow, we have studied the effect
of the aspect ratio $AR$. The mass center of an elliptic cylinder also stays at the 
middle between two walls if it is placed there initially.  Since the elliptic 
cylinder has zero angular velocity when the particle Reynolds number is larger than 
the critical value $Re_{p,c}$, the behavior of the rotation of an elliptic cylinder 
is different from that of the circular cylinder. In Fig. \ref{fig:8},
the minimal angular velocity decreases rapidly to about zero when
the particle Reynolds number $Re_p$ is closer but less than $Re_{p,c}$ 
since the motion of the elliptic cylinder is about to transit into the one with a 
fixed orientation in linear shear flow.  The minimal angular velocity decreases faster
for the smaller aspect ratio shows that the slender shape is easier to reach a stable 
orientation in linear shear flow.
But its maximal angular velocity shown in Fig. \ref{fig:8} have 
different behavior since the maximal angular velocity happens when the direction of 
the long axis is about perpendicular to the shear direction.
For the small particle Reynolds numbers, the maximal and minimal values of the angular velocity
are close to the Jeffery's solution as in Fig. \ref{fig:8}.

\vskip 4ex
\noindent{\bf 4.3. The equilibrium position}
\vskip 2ex

\begin{figure}
\begin{center}
\leavevmode
\epsfxsize=5.in
\epsffile{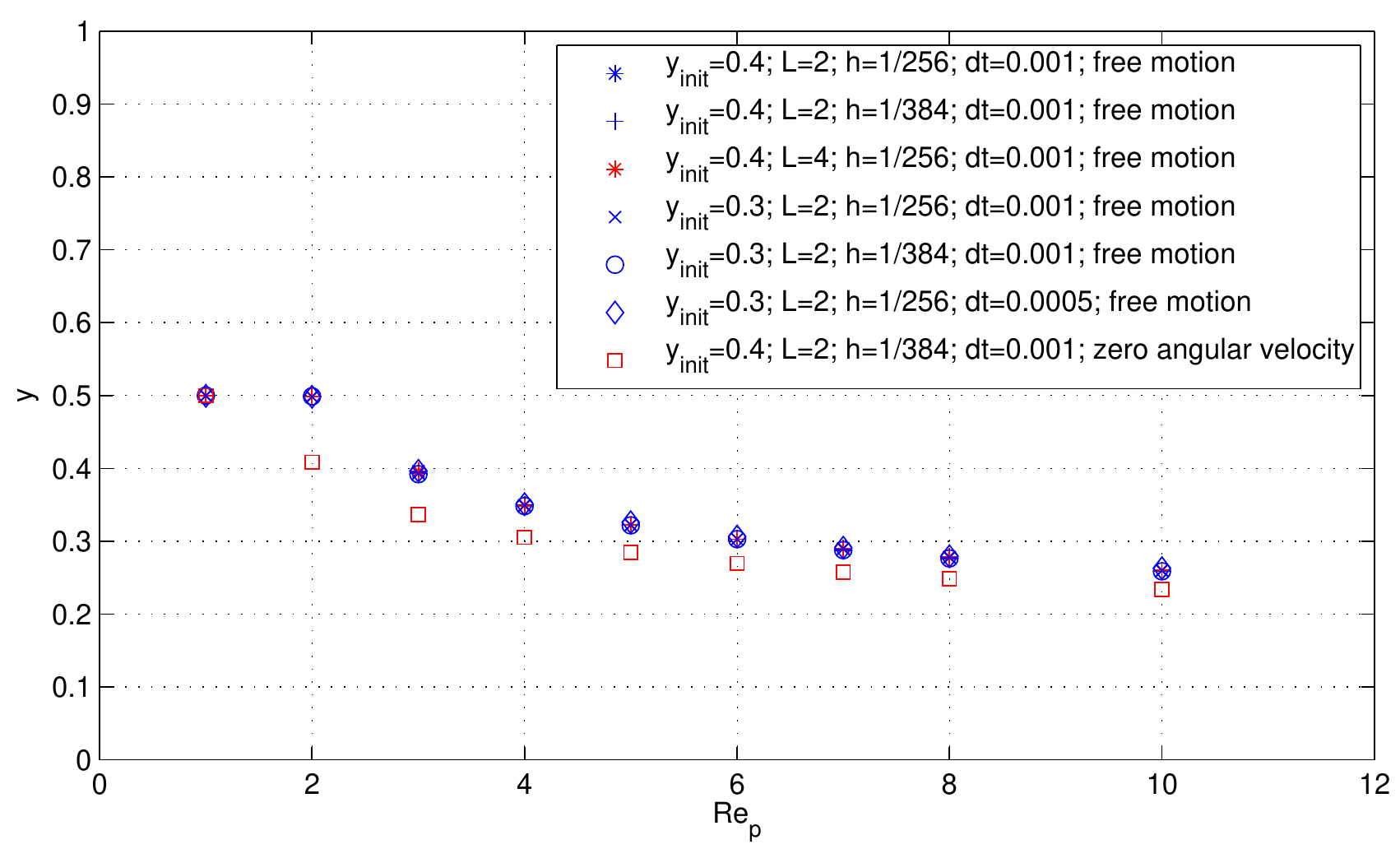} 
\end{center}
\caption{The equilibrium height of the mass center of a circular cylinder versus $Re_p$ for $\kappa=0.25$.} \label{fig:9}
\end{figure}

\begin{figure}
\begin{center}
\leavevmode
\epsfxsize=5.in
\epsffile{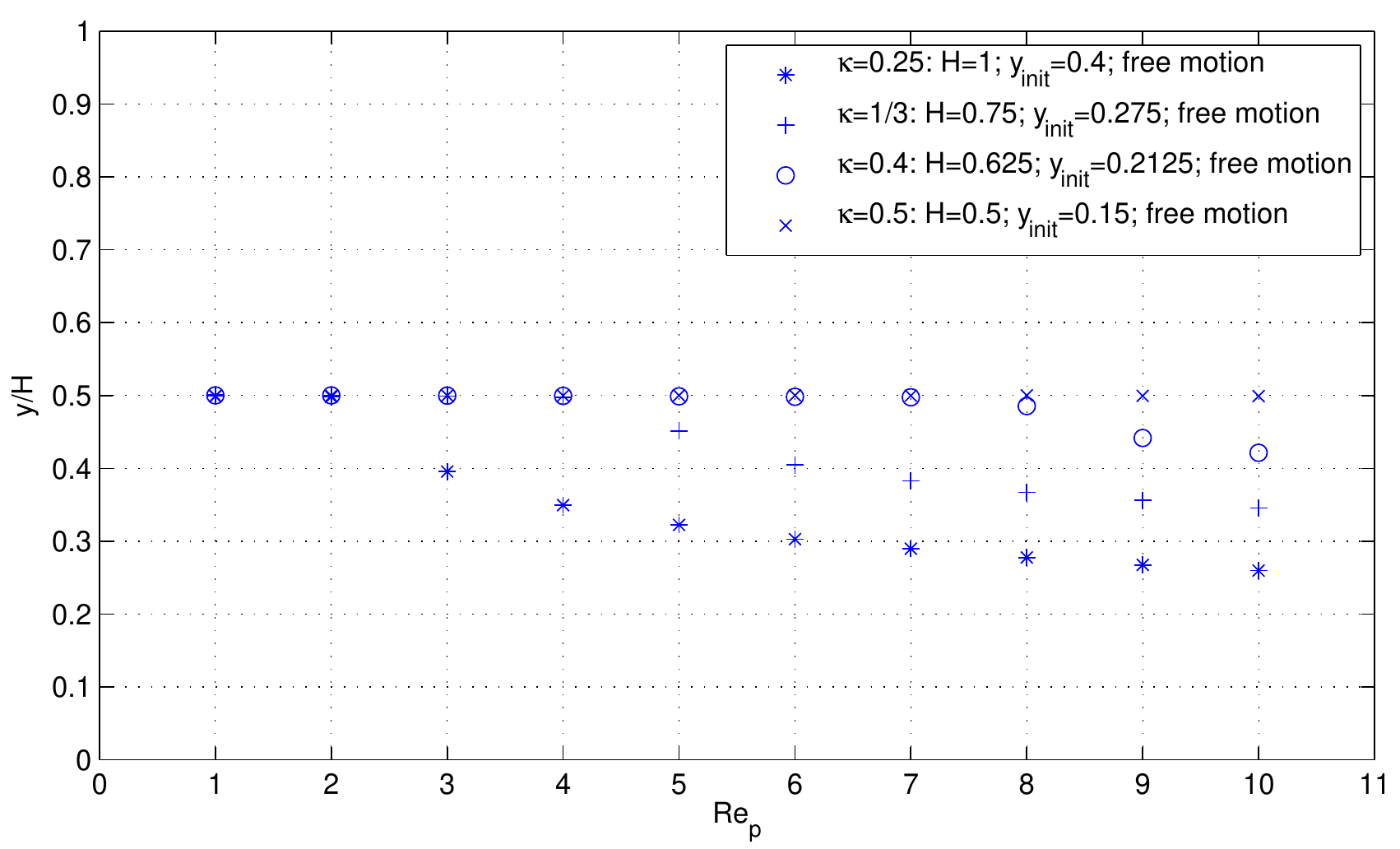} 
\end{center}
\caption{The equilibrium height of the mass center of a circular cylinder versus $Re_p$ for various values
of $\kappa$.} \label{fig:10}
\end{figure}

\begin{figure}
\begin{center}
\leavevmode
\epsfxsize=4.in
\epsffile{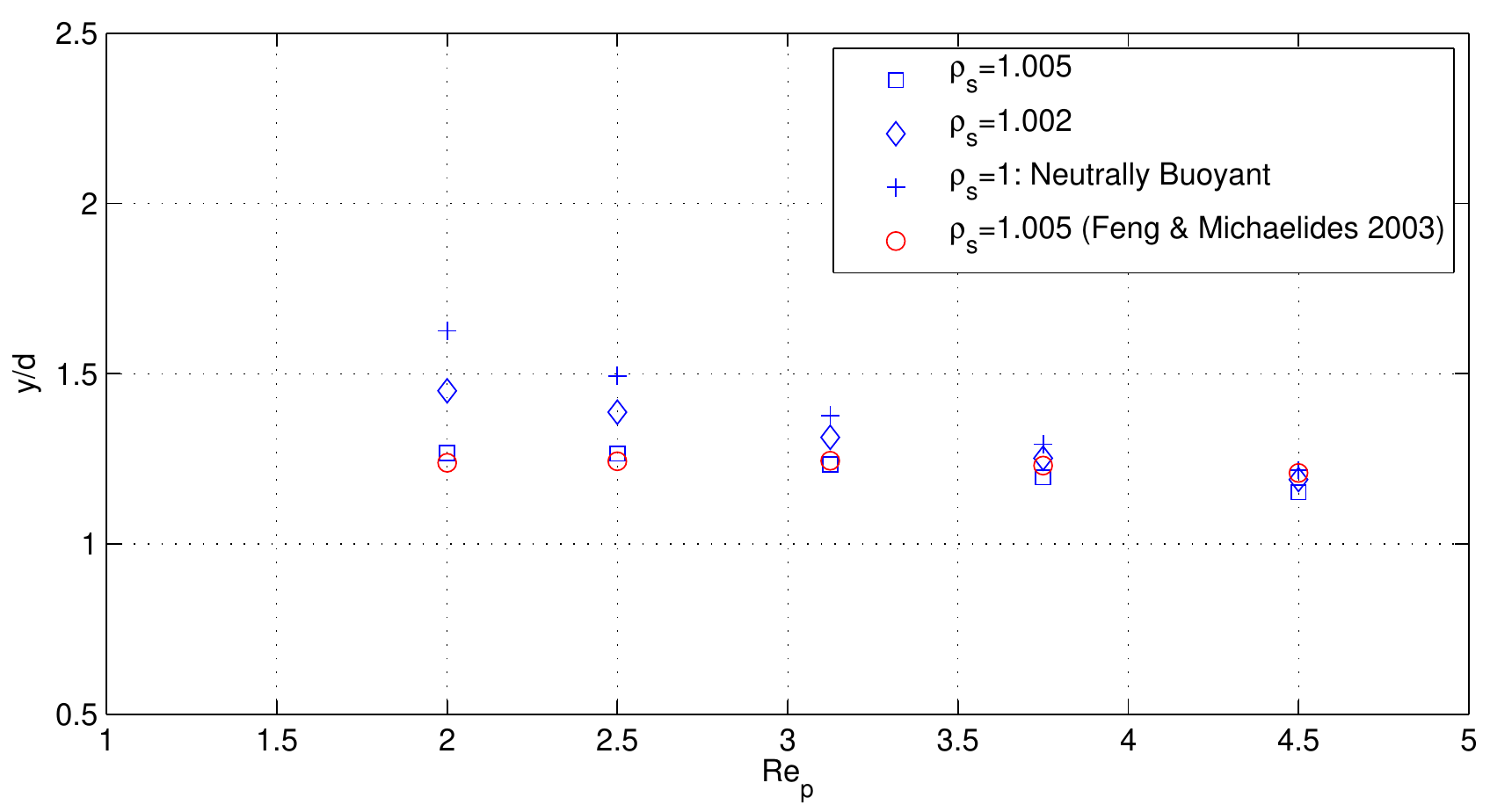} 
\end{center}
\caption{The comparison of the equilibrium height of the mass center
of a non-neutrally buoyant circular cylinder  and that of a neutrally buoyant circular cylinder versus $Re_p$.} \label{fig:11}
\end{figure}

In Ho and Leal \cite{Ho1974} and Vasseur and Cox \cite{Vasseur1976}, they concluded that the 
sphere reaches a stable lateral equilibrium position which is the midway between the walls
for small particle Reynolds numbers. 
In Feng et al. \cite{FengJ1994}, the cylinder migrates back to the midway between two walls 
at $Re_p=0.625$ when placing it away from the middle between two walls. Feng et al. have suggested 
that that three factors, namely the wall repulsion due to a lubrication effect, the slip velocity, 
and the Magnus type of lift, are possible responsible for the lateral migration. 
In the previous section, we have obtained that
the centerline is always (at least in the range we have studied in this paper) 
the equilibrium position for the cylinder of elliptic or circular
shape when it is positioned there initially. When placing the mass center initially away 
from the middle between two walls, the mass center may not migrate back to the centerline. 

\begin{figure}[ht]
\begin{center}
\leavevmode
\epsfxsize=3.7in
\epsffile{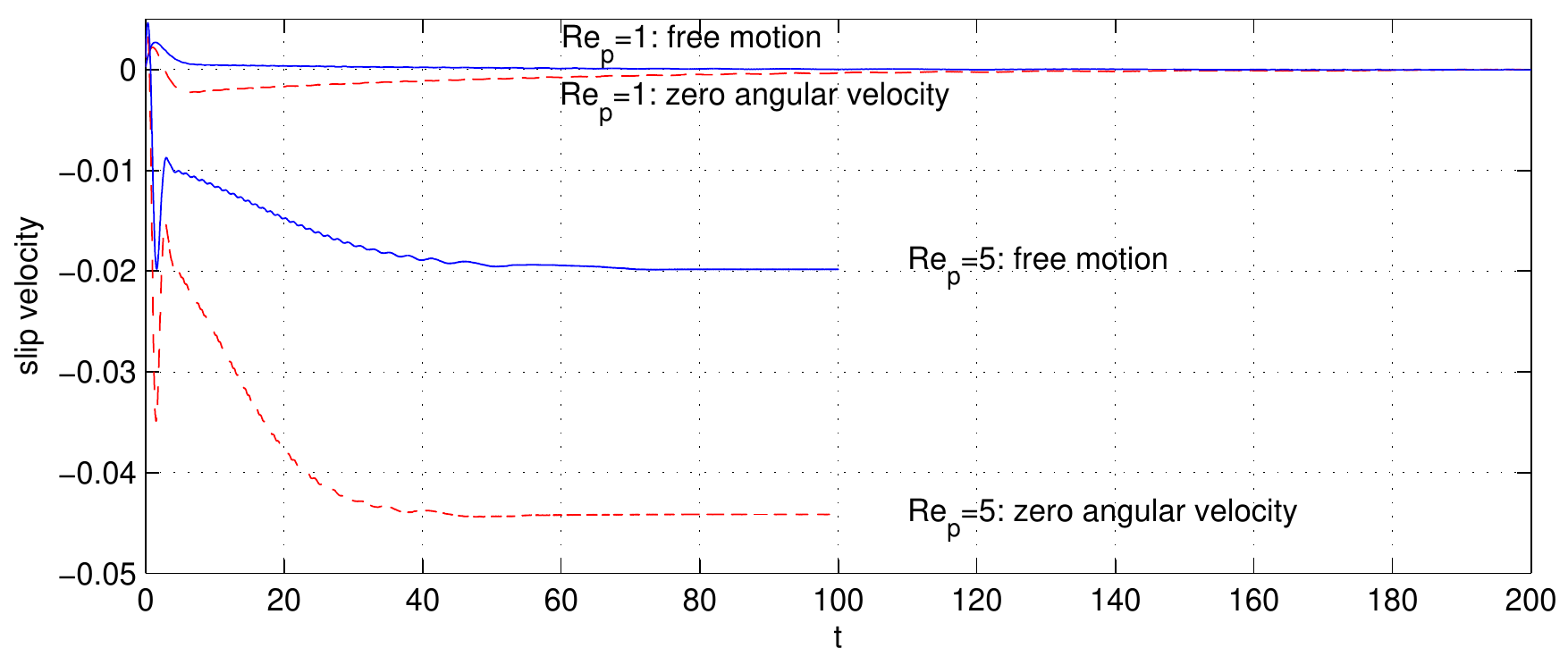}\\
\vskip 1ex \hskip 5pt
\epsfxsize=3.7in
\epsffile{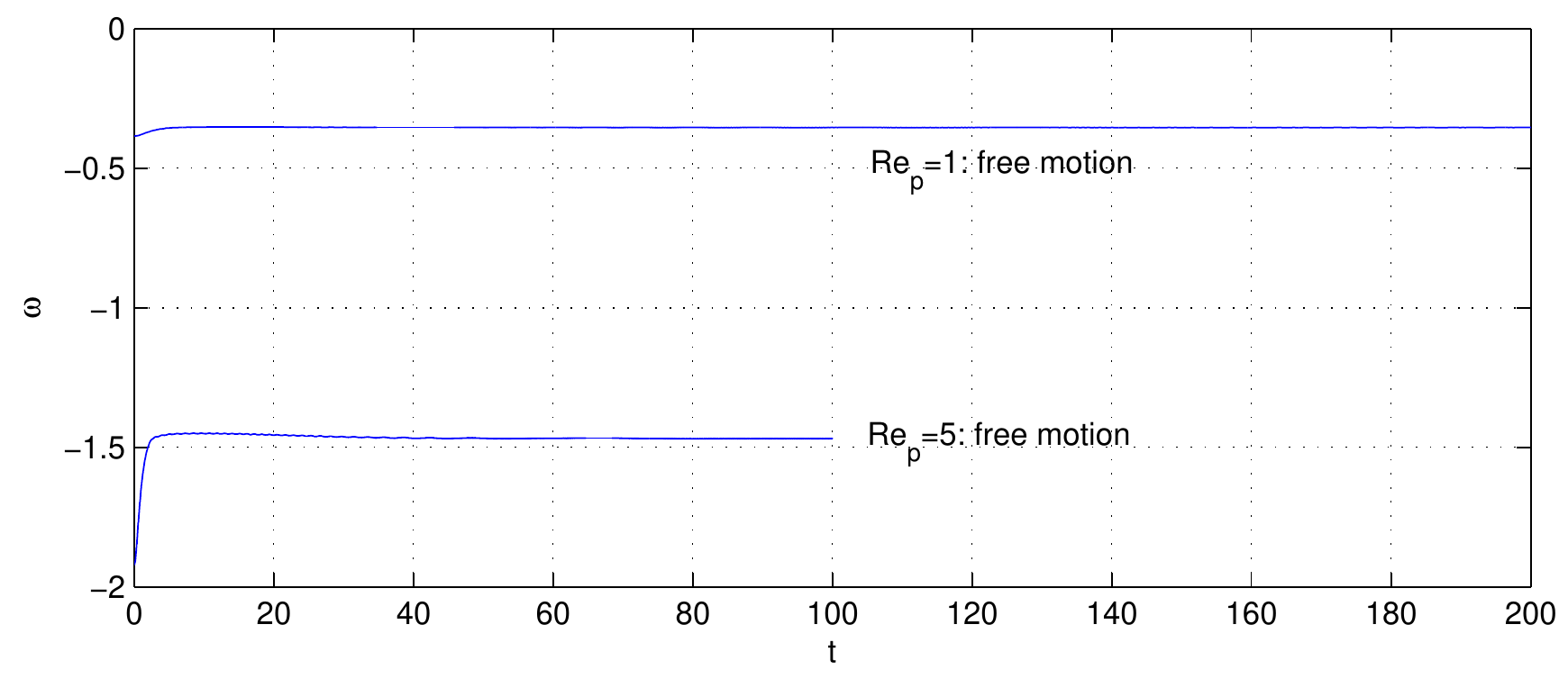}\\
\vskip 1ex \hskip 5pt
\epsfxsize=3.65in
\epsffile{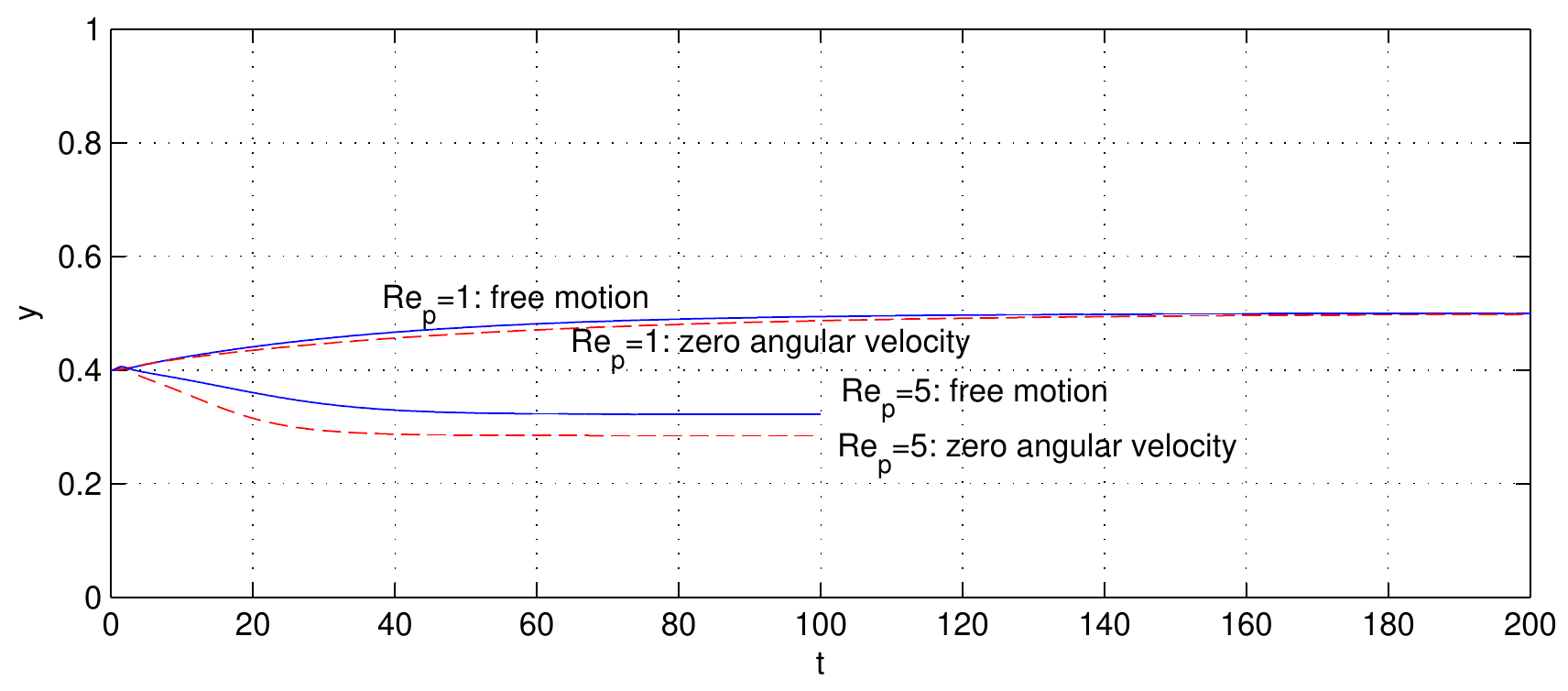}
\end{center}
\caption{The histories of the slip velocity (top), the angular velocity (middle), 
and the height of the mass center (bottom) of the cylinder for $Re_p=$1 and 5:  
free motion (solid blue line) and zero angular velocity (red dashed line). } \label{fig:12}
\end{figure}

For the cases of a circular cylinder of the confined ratio $\kappa=0.25$ as in the previous section, 
the radius of the circular cylinder is $a=0.125$, the density of the fluid 
is $\rho=1$, and the kinetic viscosity is $\nu=0.012$. To check the effect of the length $L$ of the 
channel, the initial height $y_{init}$, the mesh size $h$ and time step $dt$ on the equilibrium height
of the mass center, we have considered different sets of parameter values as indicated in Fig. \ref{fig:9}.
The final equilibrium heights in Fig. \ref{fig:9} are almost the same for each value of $Re_p$
considered here except those with zero angular velocity constraint. 
When $Re_p=$1 and 2, the mass center of the freely moving cylinder migrates 
back to the middle between two walls.  But for higher particle Reynolds numbers, we have obtained 
an equilibrium height which is between the centerline and the wall.  
To find out the effect of the walls  on the final equilibrium height,
we have varied the distance $H$ between two walls. The initial height $y_{init}$ 
is always 0.1 unit below the centerline. The length of the computational domain is $L=2$.
The equilibrium height versus the particle Reynolds number is shown in Fig. \ref{fig:10}. 
As expected, the wall repulsion force is stronger for the larger confined ratio. 
Hence the cylinder migrates back to the midway between two walls for $\kappa=0.5$
for the range of $Re_p$ considered here and 
the critical particle Reynolds number $Re_{p,c}$ is increasing when increasing the confined ratio.
We believe that the symmetric breaking shown in Figs. \ref{fig:9} and  \ref{fig:10} is not 
a numerical artifact. To further validate this phenomenon, we have compared ours with 
the results obtained in Feng and Michaelides \cite{Feng2003} where only the non-neutrally circular cylinders
were considered. For the results presented in Fig. \ref{fig:11},  the radius of circular cylinder is
$a=0.05$, the initial height is $y_{init}=0.1$, the kinetic viscosity is $\nu=0.05$, and the length and 
the height of the computation domain are $L=2$ and $H=1$, respectively. The density $\rho_s$
of the non-neutrally circular cylinder is either 1.002 or 1.005. We have applied the DLM/FD 
method developed in, e.g., \cite{RG1999, RG2001} to obtain the computational results for the  
cases of the non-neutrally circular cylinder. Our results are in a good agreement with 
the results obtained in Feng and Michaelides \cite{Feng2003}. The transition 
of the equilibrium height from the one associated with a non-neutrally buoyant cylinder to that of
a neutrally buoyant cylinder is consistent and support the existence of the symmetric breaking of 
the equilibrium height.

\begin{figure}[t]
\begin{center}
\leavevmode
\epsfxsize=5.in
\epsffile{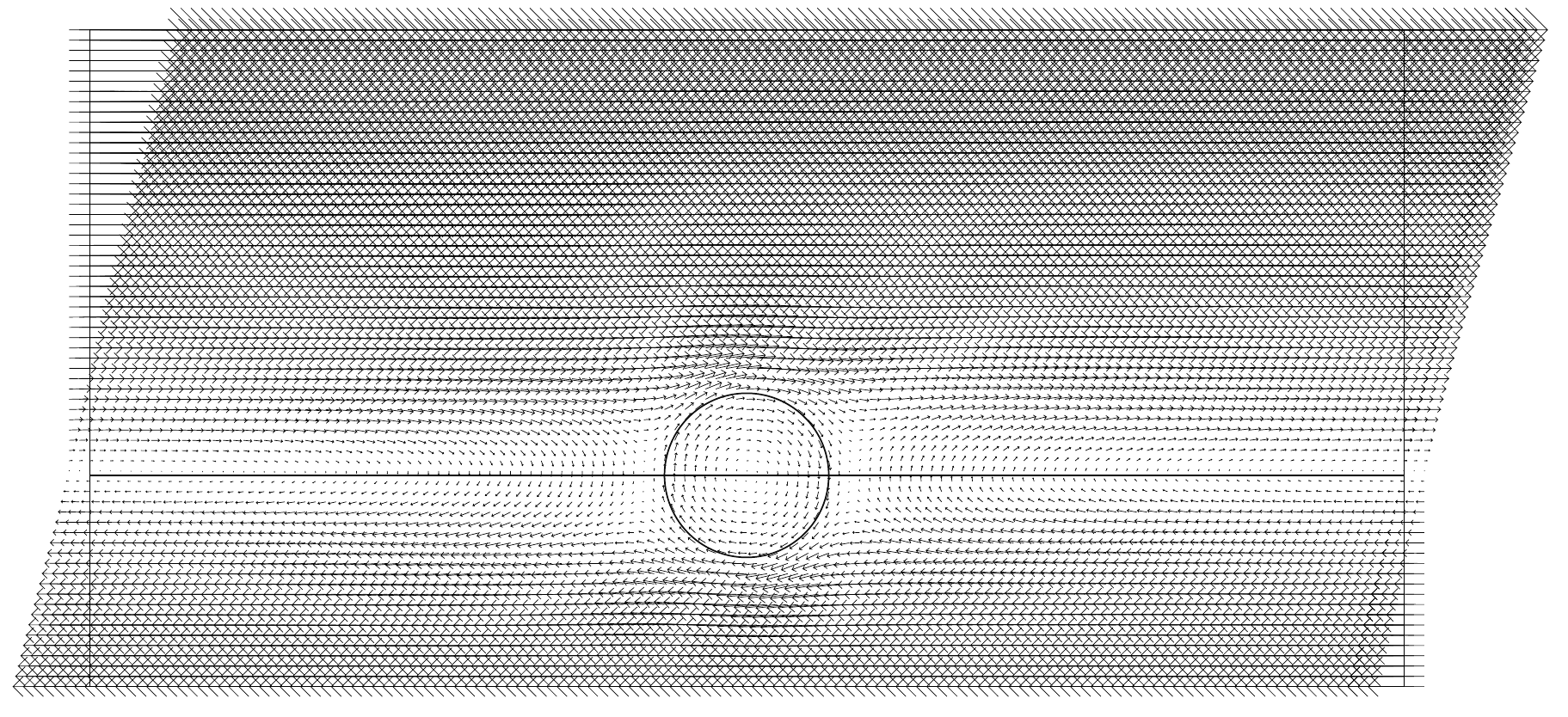}\\
\vskip 1ex
\epsfxsize=5.in
\epsffile{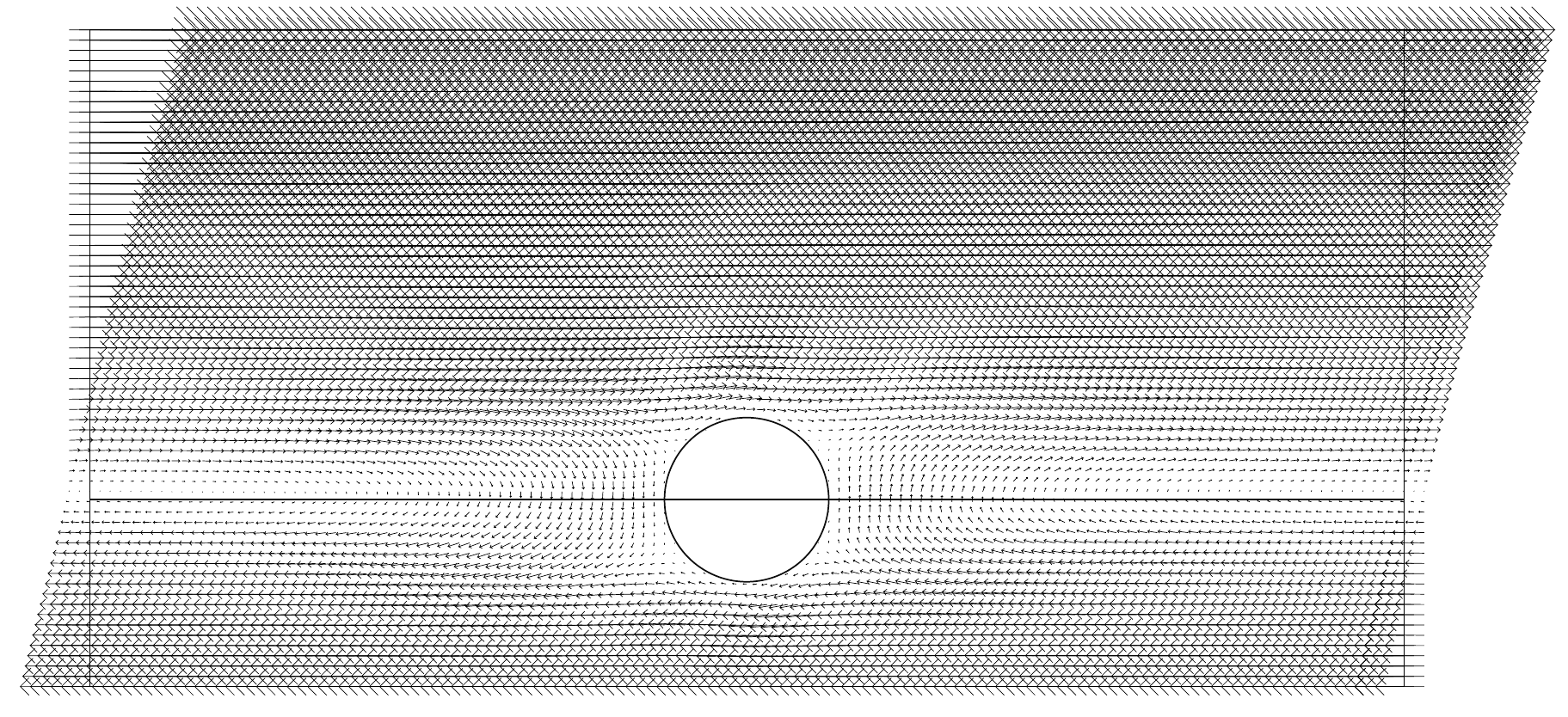}
\end{center}
\caption{A snapshot of the velocity field of $Re_p=5$ obtained by following the cylinder
mass center which is moving to the left: freely motion (top) and zero angular velocity (bottom).} \label{fig:13}
\end{figure}

For the effect of the Magnus lift associated with the rotation, we
have considered the cases of a circular cylinder of $\kappa=$0.25 moving with zero 
angular velocity. The equilibrium height of non rotating cylinder is lower than the one of the freely 
moving cylinder for $Re_p \ge 2$ as in Fig. \ref{fig:9}. These results indicate 
that the Magnus lift does play a significant role as expected. The similar results concerning
the Magnus lift have also been observed in Feng and Michaelides \cite{Feng2003}. When $Re_p$=1, the
wall repulsion force is strong enough to push the cylinder back to the centerline even without 
the help from the Magnus lift. 

We now focus on the effect of the slip velocity. Since the initial position of the cylinder 
is below the centerline, it is moving to the left in the lower region of the computational
domain due to the given boundary condition (see Fig. \ref{fig:1}). 
For getting the slip velocity, we first compute the fluid horizontal speed
on the streamline through the mass center in front of the cylinder at the distance of 
the half of the computational domain width and then minus the horizontal speed of the disk
to obtain the slip velocity. The negative slip velocity means that the cylinder
speed to the left is slower than the fluid speed to the left since the both signs are negative. 
We then have that the cylinder lags the fluid.
For $Re_p$=1, the slip velocity of the non rotating cylinder becomes negative after a short initial 
transition period as in Fig. \ref{fig:12}. The one associated with free moving circular cylinder
remains positive for, at least, the first 110 time units and then oscillates about zero.  
The cylinder with no rotation lags the fluid. But both cylinders migrate back to the middle
between two walls due to that the wall repulsion force dominates the very weak slip velocity 
effect. We can see freely moving circular cylinder moves toward the centerline faster than 
the one with no rotation does in Fig. \ref{fig:12}. 
When $Re_p$=5, the slip velocity becomes negative after the initial transition period  
as in Fig. \ref{fig:12}. The cylinder for the both cases lags the fluid. The slip velocity of the 
one with no rotation  is about two times larger than the other one.  The rotating velocity of each case
of free motion is about constant speed after a short initial transition period.  
Fig. \ref{fig:13}  shows that the relative velocity field to the horizontal velocity of
the cylinder mass center in which we can clearly see that the cylinder lags the fluid. Both cases
have stronger slip velocity which creates force pushing the cylinder to the region with faster 
flow speed, which is the region next to the bottom wall. Without the extra help from the Magnus lift, 
the one without rotation is closer to the wall. Hence when the initial position is not at the middle 
of two walls, the balance between the effect of the slip velocity, the Magnus lift 
and the wall repulsion does play a role for determining the equilibrium position of the cylinder.

\vskip 3ex
\noindent{\large\bf 5. Conclusion}
\vskip 2ex
We have investigated the motion of a neutrally
buoyant cylinder of circular or elliptic shape in two dimensional shear flow 
of a Newtonian fluid by direct numerical simulation. 
The numerical results are validated by comparisons with 
existing theoretical, experimental, and numerical results, including a power law of the 
normalized angular speed versus the particle Reynolds number. 
The rapid slow down of the normalized minimal angular speed of an elliptic cylinder
is totally different from the behavior of the circular cylinder since the motion 
of an elliptic cylinder can transit  from rotating to a fixed orientation 
when the particle Reynolds number is increased beyond the critical 
value. The midway between two walls is an expected equilibrium position of 
the cylinder mass center in shear flow. But when placing the particle 
away from the centerline initially, it migrates toward another equilibrium 
position between the wall and the centerline for higher Reynolds numbers which 
is caused by the interplay between the slip velocity, the Magnus force, and the 
wall repulsion force. The further study of the motion of a cylinder of axisymmetric 
shape in simple shear flow will be reported in a forthcoming paper.

\vskip 2ex
\noindent{\large\bf Acknowledgments.} T.-W. Pan acknowledges the support 
by the US NSF under Grant No. DMS-0914788. 
S.-L. Huang, S.-D. Chen, C.-C. Chu, C.-C. Chang acknowledge the support by 
the National Science Council (Taiwan, ROC) under Contract Numbers, 
NSC97-2221-E-002-223-MY3, NSC99-2628-M-002-003 and NSC100-2221-E-002-152-MY3.

\end{document}